\documentclass[fleqn,usenatbib]{mnras}

\usepackage{newtxtext,newtxmath}

\usepackage[T1]{fontenc}
\usepackage{longtable}
\usepackage{subfigure}
\usepackage{lscape}
\DeclareRobustCommand{\VAN}[3]{#2}
\let\VANthebibliography\thebibliography
\def\thebibliography{\DeclareRobustCommand{\VAN}[3]{##3}\VANthebibliography}

\def\EE{$E_{\mathrm{iso}}$-$E_{\mathrm{p}}$}
\usepackage{color}

\usepackage{graphicx}	
\usepackage{amsmath}	

\title[$E_{\mathrm{iso}}$-$E_{\mathrm{p}}$ correlation: calibration and applications]{$E_{\mathrm{iso}}$-$E_{\mathrm{p}}$ correlation of gamma ray bursts: calibration and cosmological applications}

\author[X. D. Jia et al.]{
X. D. Jia,$^{1}$
J. P. Hu,$^{1}$
J.Yang$^{1}$
B. B. Zhang$^{1,2}$
and F. Y. Wang$^{1,2}$\thanks{E-mail: fayinwang@nju.edu.cn}
\\
$^{1}$School of Astronomy and Space Science, Nanjing University, Nanjing 210093, China\\
$^{2}$Key Laboratory of Modern Astronomy and Astrophysics (Nanjing University), Ministry of Education, Nanjing 210093, China\\
}

\date{Accepted 2022 August 18. Received 2022 August 18; in original form 2022 June 21}

\pubyear{2022}

\begin{document}
\label{firstpage}
\pagerange{\pageref{firstpage}--\pageref{lastpage}}
\maketitle

\begin{abstract}
Gamma-ray bursts (GRBs) are the most explosive phenomena and can be used to study the expansion of Universe. In this paper, we compile a long GRB sample for the $E_{\mathrm{iso}}$-$E_{\mathrm{p}}$ correlation from Swift and Fermi observations. The sample contains 221 long GRBs with redshifts from 0.03 to 8.20. From the analysis of data in different redshift intervals, we find no statistically significant evidence for the redshift evolution of this correlation. Then we calibrate the correlation in six sub-samples and use the calibrated one to constrain cosmological parameters. Employing a piece-wise approach, we study the redshift evolution of dark energy equation of state (EOS), and find that the EOS tends to be oscillating at low redshift, but consistent with $-1$ at high redshift. It hints a dynamical dark energy at $2\sigma$ confidence level at low redshift.
\end{abstract}

\begin{keywords}
dark energy -- cosmological parameters -- gamma-ray burst: general 
\end{keywords}


\section{Introduction}\label{sec:intro}
The study of type Ia supernovae (SNe Ia) revealed the evidence of accelerating expansion of the universe \citep{1998AJ....116.1009R,1999ApJ...517..565P}, which shed light on the mysterious component — dark energy. Additionally, several independent observations have confirmed the accelerated expansion of the universe, including the cosmic microwave background \citep[CMB;][]{2003ApJS..148..175S}, and the baryonic acoustic oscillations \citep[BAO;][]{2005ApJ...633..560E}. The $\Lambda$CDM model successfully accounts for most cosmological observations \citep{2020A&A...641A...6P,2021MNRAS.504.2535I,2022MNRAS.513.5686C}. However, other dark energy models can not be ruled out due to the precision of current measurements. Currently, the highest redshift of SNe Ia is 2.26 \citep{2018ApJ...859..101S} and there is still blankness between SNe Ia and CMB. Fortunately, high redshift observations (for example GRBs and quasars) provide an opportunity for us to explore the cosmic blank history.
\par
GRBs are the most violent phenomena in the Universe, which have the isotropic equivalent energy up to $10^{54}$ erg \cite[for reviews, see][]{2009ARA&A..47..567G, 2015PhR...561....1K}. GRBs are usually classified into two types based on the duration time ($T_{90}$): long GRBs ($T_{90} > 2 s$) and short GRBs ($T_{90} < 2 s$) \citep{1993ApJ...413L.101K}. The former is thought to result from the core collapse of massive stars $\left(\geq 25 M_{\odot}\right)$. The progenitor of the latter is thought to be mergers of compact object binary \citep{2009ARA&A..47..567G,2017ApJ...848L..13A}. The redshift range that they cover is very wide, up to $z\sim$ 9.40, making them as attractive cosmological probes \citep{2015NewAR..67....1W}. Hence, there have been a lot of studies demonstrating that GRBs are useful in extending the Hubble diagram to high redshifts
\citep{2001ApJ...562L..55F,2004ApJ...612L.101D,2004ApJ...616..331G,2005ApJ...633..611L,2007ApJ...660...16S,2012A&A...543A..91W}. To use GRBs as "standard candles", researchers have found several correlations between various characteristics of the prompt emission and the afterglow emission \citep{2002A&A...390...81A,2004ApJ...616..331G,2005ApJ...633..603X,2006MNRAS.369L..37L}. Attempts to use GRBs for constraining cosmological parameters have also obtained encouraging results \citep{2009MNRAS.400..775C,2014ApJ...783..126P,2019MNRAS.486L..46A,2019ApJS..245....1T,2021MNRAS.501.1520C,2021ApJ...914L..40D,2021MNRAS.507..730H,2021JCAP...09..042K,2021ApJ...920..135X,2022arXiv220408710C,2022MNRAS.510.2928C,2022MNRAS.512..439C,2022MNRAS.514.1828D}. Some reviews on luminosity correlations and cosmological applications of GRBs can be found in \cite{2015NewAR..67....1W,2017NewAR..77...23D,2018AdAst2018E...1D,2018PASP..130e1001D}.

In this paper, we adopt the $E_{\mathrm{iso}}-E_{\mathrm{p}}$ correlation to explore the high-redshift universe using a long GRB sample from Swift and Fermi catalogs. The $E_{\mathrm{iso}}-E_{\mathrm{p}}$ correlation that the isotropic energy $E_{\mathrm{iso}}$ is correlated with the rest-frame peak energy $E_{\mathrm{p}}$ was discovered by \cite{2002A&A...390...81A} with a small sample of GRBs. Subsequently, \cite{2016A&A...585A..68W} updated 42 long GRBs and calibrated the $E_{\mathrm{iso}}-E_{\mathrm{p}}$ correlation with SNe Ia. The combination of GRBs and SNe Ia gave $\Omega_{m}=0.271\pm0.019$ and $H_0=70.1\pm0.2$ km s$^{-1}$ Mpc$^{-1}$ for the flat $\Lambda$CDM model. Recently, through analysing the correlation parameters and six different cosmological models simultaneously, \cite{2020MNRAS.499..391K} found that the $E_{\mathrm{iso}}-E_{\mathrm{p}}$ correlation is independent of cosmological models but GRB data can not constrain cosmological parameters to a great extent at present. In order to constrain cosmological model parameters strictly, the $E_{\mathrm{iso}}-E_{\mathrm{p}}$ correlation is also capable of being combined with the Combo-relation. The results are consistent with flat $\Lambda$CDM model, dynamical dark energy models and non-spatially-flat models \citep{2021JCAP...09..042K}. The data of Observational Hubble Dataset measurements (OHD) also help to constrain the cosmological parameters. \cite{2021MNRAS.503.4581L} calibrated the $E_{\mathrm{iso}}-E_{\mathrm{p}}$ correlation with the data of OHD and generated mock catalogs with machine learning techniques. They tested the $\Lambda$CDM model and the Chevallier-Polarski-Linder parametrization, finding possible extensions of the $\Lambda$CDM model toward a weakly evolving dark energy evolution. Combining GRBs with other probes, a joint analysis of the $H(z)$+BAO+quasar+HII starburst galaxy+GRBs data provides $\Omega_{m} = 0.313\pm0.013$ in a model-independent way \citep{2021MNRAS.501.1520C}. Their results provide a supporting consistency for the $\Lambda$CDM model, but it could not rule out mild dark energy dynamics. \cite{2022arXiv220700440L} used OHD and BAO data to calibrate the $E_{\mathrm{iso}}-E_{\mathrm{p}}$ correlation. Basing on the assumption that the GRB data obey a special redshift distribution, \cite{2022ApJ...935....7L} constrained $\Omega_m$ to be $0.308^{+0.066}_{-0.230}$ and $0.307^{+0.057}_{-0.290}$ with an improved $E_{\mathrm{iso}}-E_{\mathrm{p}}$ correlation in the $\Lambda$CDM model and $w$CDM model, respectively.
\par
In order to calibrate the correlations of GRBs, many methods have been tried. Using Bézier parametric curve to approximate the Hubble function is a model independent calibration method. \cite{2019MNRAS.486L..46A} fitted the $E_{\mathrm{iso}}-E_{\mathrm{p}}$ correlation with 193 long GRBs and the results show that the $\Lambda$CDM model is statistically superior to the $w$CDM model. The slope parameter of the Combo-relation was calibrated from small sub-samples of GRBs lying almost at the same redshift. And the intercept parameter was determined from the SNe Ia located near the GRBs \citep{2021ApJ...908..181M}. Another method is using the Gaussian process with the data of OHD to calibrate GRB correlations \citep{2022ApJ...924...97W}. Considering the number of the GRB sample used in this paper, we decide to study the $E_{\mathrm{iso}}-E_{\mathrm{p}}$ correlation by dividing them into several sub-samples.
\par
Increasing GRB observations have given rise to use the $E_{\mathrm {iso}}-E_{\mathrm{p}}$ correlation in cosmology. In this study, we use 221 GRBs to test the $E_{\mathrm {iso}}-E_{\mathrm{p}}$ correlation. The full sample is based on \cite{2016A&A...585A..68W}, and 29 GRBs from \cite{2019MNRAS.486L..46A}, and 49 GRBs from Fermi catalog are added. The spectral parameters are also taken from Fermi catalog. After converting the observed values to the cosmological rest frame, the bolometric fluence is calculated with the $k$-correction. For the $E_{\mathrm {iso}}-E_{\mathrm{p}}$ correlation, in view of the extrinsic scatter $\sigma_{\mathrm{ext}}$ should also depend on hidden variables, we take $\sigma_{\mathrm{ext}}$ assigned to $E_{\mathrm{iso}}$. This is consistent with the method proposed by \cite{2005physics..11182D} and more detail are discussed in Sec. \ref{Sec3}. The possible redshift evolution is studied by dividing the full sample into five redshift bins. The results show that the correlation does not have an evolution with redshift within 2$\sigma$ confidence level. To avoid the circularity problem, six groups within small redshift ranges are selected from the full GRB sample. The redshift range is small so that the $E_{\mathrm {iso}}-E_{\mathrm{p}}$ correlation in each sub-sample is almost model-independent. The correlation can be calibrated. 
\par
This paper is structured as follows. In Sec. \ref{Sec2}, we introduce the GRB sample and perform the $k$-correction. In Sec. \ref{Sec3}, we fit coefficients of the $E_{\mathrm {iso}}-E_{\mathrm{p}}$ correlation, and test whether the correlation evolves with redshifts. To avoid the circularity problem, we calibrate the correlation in sub-samples. In Sec. \ref{Sec4}, we use the calibrated correlation to constrain cosmological parameters. In Sec. \ref{Sec5}, we study the dark energy EOS in a model-independent way. We summarize the results and make some discussions in Sec. \ref{Sec6}.

\section{GRBs sample}\label{Sec2}
The Swift satellite has provided a large number of GRBs with redshifts. Its three instruments give scientists the ability to scrutinize GRBs. But the BAT instrument of this satellite is only capable of detecting energies up to 150 keV \citep{Gehrels2004}, which is lower than the average peak energy of GRBs \citep{2006ApJS..166..298K}. Hence, for many GRBs observed by the Swift satellite, the fluence and $E_{\mathrm{p,obs}}$ can not be directly determined. While the Fermi satellite has two main instruments: the Large Area Telescope (LAT) and the Gamma-ray Burst Monitor (GBM). It studies the cosmos between the energy range of 10 keV to 300 GeV. The most significant advantage is that Fermi is able to determine all the spectral parameters in the Band function. Consequently, we compile a sample of long GRBs that appear in both Swift and Fermi catalogs.
\par
Basing on the data set constructed by \cite{2016A&A...585A..68W}, we collect all GRBs with information of fluence, peak energy, and power law index from Fermi catalogue including observations from August 2008 to June 2021 \citep{2014ApJS..211...12G,2014ApJS..211...13V,2016ApJS..223...28N,2020ApJ...893...46V}. The redshifts are obtained from the Swift database\footnote{\url{https://swift.gsfc.nasa.gov/archive/grb_table.html/}}. Noting that some of the GRBs listed in the Fermi catalogue present no values for the spectral parameters or $E_{\mathrm{p,obs}}$. We download the corresponding time-tagged event dataset from Fermi public data archive\footnote{\url{https://heasarc.gsfc.nasa.gov/FTP/fermi/data/gbm/daily/}}. Data reduction and analysis follow the procedures discussed by \cite{2011ApJ...730..141Z,2016ApJ...816...72Z}. We select up to three sodium iodide (NaI) detectors and one bismuth germanium oxide (BGO) detector based on the method proposed by \cite{2021ApJ...923L..30Z} for all GRBs to perform the spectral fitting. Meanwhile, to ensure sufficient detector response, the viewing angles from the GRB location should be less than 60 degrees for NAI detectors and closest for BGO detector. For each detector, the source spectrum and background spectrum in a specific time interval are generated by summing the total and background photons in each energy channel, respectively. And the response matrices are required using the GBM Response Generator\footnote{\url{https://fermi.gsfc.nasa.gov/ssc/data/analysis/rmfit/gbmrsp-2.0.10.tar.bz2}}. Then we use McSpecFit discussed by \cite{2018NatAs...2...69Z} to perform the spectral fitting, which packages the nested sampler Multinest and utilizes pastat as the statistic to constrain parameters. The band function is employed to fit spectra.
\par
The GRBs prompt emission spectrum can be described as an empirical spectral function, which is a broken power law known as the Band function \citep{1993ApJ...413..281B}
\begin{equation}
\Phi(E)= \begin{cases}A E^{\alpha} \mathrm{e}^{-(2+\alpha) E / E_{\mathrm{p}, \mathrm{obs}}} & \mathrm { if \ } E \leq \frac{\alpha-\beta}{2+\alpha} E_{\mathrm{p}, \mathrm{obs}} \\ B E^{\beta} & \mathrm { otherwise, }\end{cases}    
\end{equation}
where $E_{\mathrm{p}, \mathrm{obs}}$ is the observed peak energy , $\alpha$ and $\beta$ are the low- and high-energy indices, respectively. With $E_{\mathrm{p}, \mathrm{obs}}$ and redshift $z$, we get the peak energy in the rest frame by $E_{\mathrm{p}}=E_{\mathrm{p}, \mathrm{obs}} \times(1+z)$. 
\par
The bolometric fluence is calculated in the energy band of $1-10^{4} \mathrm{keV}$ by $k$-correction \citep{2001AJ....121.2879B}
\begin{equation}
S_{\mathrm {bolo }}=S \times \frac{\int_{1 /(1+z)}^{10^{4} /(1+z)} E \Phi(E) \mathrm{d} E}{\int_{E_{\min }}^{E_{\max }} E \Phi(E) \mathrm{d} E},
\end{equation}
where $S$ is the observed fluence, and the detection thresholds are ($E_{\min }$, $E_{\max }$). 
\par
$E_{\mathrm{iso }}$ is the isotropic equivalent energy in gamma-ray band, which can be calculated in terms of 
\begin{equation}
E_{\mathrm{iso }}=4 \pi d_{\mathrm{L}}^{2} S_{\mathrm{bolo }}(1+z)^{-1},
\end{equation}
here $d_{\mathrm{L}}$ is the luminosity distance. The factor $(1+z)^{-1}$ transforms the duration to the source rest-frame. The luminosity distance depends on cosmological models. Here we use the standard cosmological parameters: $\Omega_{\mathrm{m}}=0.315$, $\Omega_{\Lambda}=0.685$ and $H_{0}$ = 67.4 $\mathrm{km~s}^{-1} \mathrm{Mpc}^{-1}$ \citep{2020A&A...641A...6P}, where $\Omega_{\mathrm{m}}$ is the non-relativistic matter density parameter, $\Omega_{\Lambda}$ is the cosmological constant density and $H_{0}$ is the Hubble constant. Thus the luminosity distance $d_{\mathrm{L}}$ is expressed as
\begin{equation}
\begin{aligned}
d_{\mathrm{L}}(z)=& \frac{c(1+z)}{H_{0}}\int_{0}^{z} \frac{\mathrm{d} z^{\prime}}{\sqrt{\Omega_{m}(1+z^{\prime})^{3}+\Omega_{\mathrm{\Lambda}}}}.
\end{aligned}
\end{equation}
\par
The full sample contains 221 GRBs and covers the redshift range from 0.0335 to 8.20. The GRB sample is listed in Table \ref{GRBsample}.  During the calculation, we only take into account the propagation of errors from bolometric fluence $S_{\mathrm{bolo}}$. The uncertainties from other parameters are attributed into the $\sigma_{\mathrm{ext}}$.

\section{The \EE{} correlation}\label{Sec3}
\subsection{Fitting the $E_{\mathrm{iso }}-E_{\mathrm{p}}$ correlation}
The $E_{\mathrm{iso}}-E_{\mathrm{p}}$ correlation is expressed as a logarithmic form
\begin{equation}
\log \frac{E_{\mathrm{iso }}}{\mathrm{ erg }}=a+b \log \frac{E_{\mathrm{p}}}{\mathrm{keV}}.
\end{equation}
The coefficient $a$ is the intercept parameter and $b$ is the slope parameter. 
\par
\begin{figure}
	\centering
	\includegraphics[width=0.5\textwidth,angle=0]{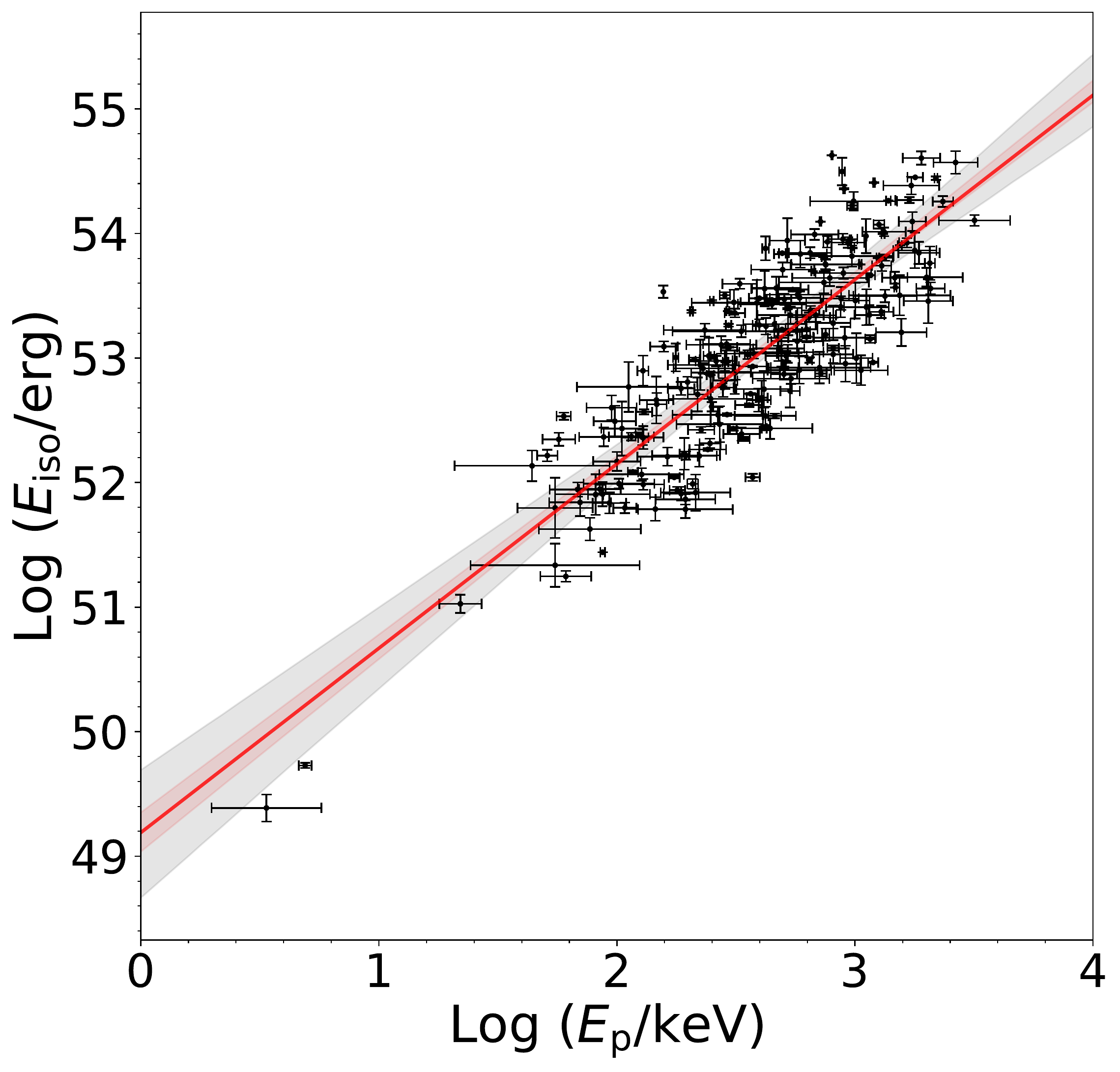}
	\caption{The $E_{\mathrm{iso }}-E_{\mathrm{p}}$ correlation with 221 long GRBs. The solid red line represents the best fit. The light red region and the light black region represent the $1 \sigma$ uncertainties and $2 \sigma$ uncertainties, respectively.}
	\label{F_EisoEp}       
    \end{figure}

As the method of fitting procedure, we use the Markov Chain Monte Carlo (MCMC) technique with the  emcee\footnote{\url{https://emcee.readthedocs.io/en/stable/}} package  to analyse our data \citep{2013PASP..125..306F}. The posterior probability density functions clearly express the best-fit values of parameters. For the fitting of the linear correlation \citep{2005physics..11182D}, the likelihood function is
    \begin{equation}
    \begin{aligned}
    \mathcal{L}\left( \Omega_{\mathrm{m}}, a, b, \sigma_{\mathrm{ext}}\right) \propto \prod_{i} &   \frac{1}{\sqrt{\sigma_{\mathrm{ext}}^{2}+\sigma_{y_{i}}^{2}+b^{2} \sigma_{x_{i}}^{2}}} \\
    & \times \exp \left[-\frac{\left(y_{i}-a-b x_{i}\right)^{2}}{2\left(\sigma_{\mathrm{ext}}^{2}+\sigma_{y_{i}}^{2}+b^{2} \sigma_{x_{i}}^{2}\right)}\right] ,
    \end{aligned}
    \end{equation}
	where $x_{i}$ and $y_{i}$ are the observational data for the $i$th GRB. Basing on the description from \cite{2005physics..11182D}, the parameter $y$ should not only depend on $x$, but also some hidden variables ($\Omega_{\mathrm{m}}$ here). Thus, we write the $E_{\mathrm {iso }}-E_{\mathrm{p}}$ correlation as $y=\log {E_{\mathrm{iso}}}/{\mathrm{erg}}$ and $x=\log {E_{\mathrm{p}}}/{\mathrm{keV}}$. The best-fit values with 1$\sigma$ uncertainties are $a=49.24\pm 0.16$, $b=1.46\pm0.06$ and $\sigma_{\mathrm{ext}}=0.39\pm0.02$, respectively.  Fig. \ref{F_EisoEp} illustrates  the $E_{\mathrm{iso }}-E_{\mathrm{p}}$ correlation for the GRB sample. 
    
	\subsection{Testing the evolution of $E_{\mathrm{iso }}-E_{\mathrm{p}}$ correlation with redshifts}
\begin{figure}
	  \centering
	  \begin{subfigure}
	     \centering
	     \includegraphics[width=0.5\textwidth,angle=0]{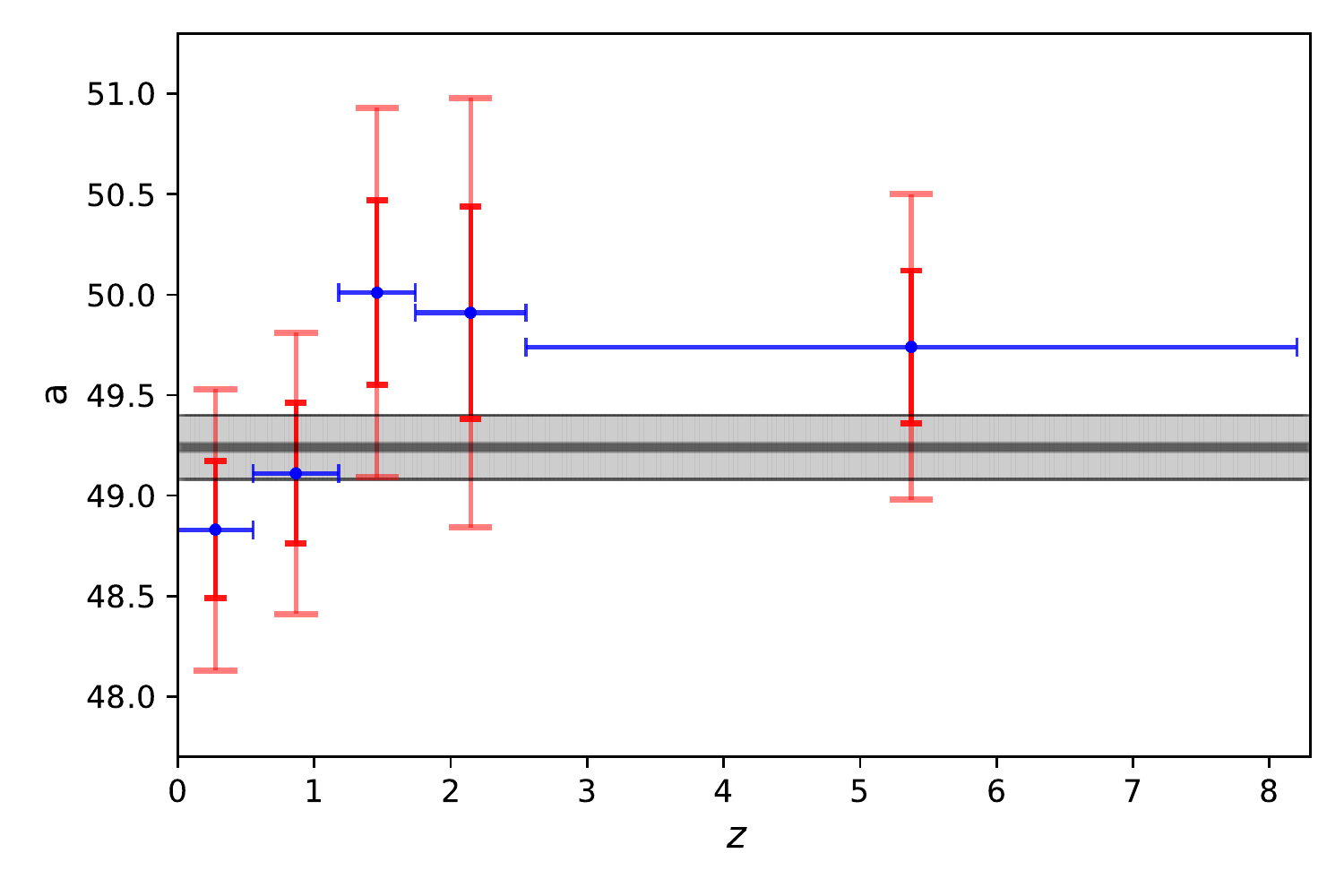}
	  \end{subfigure}
	  \centering
	  \begin{subfigure}
	     \centering
	     \includegraphics[width=0.5\textwidth,angle=0]{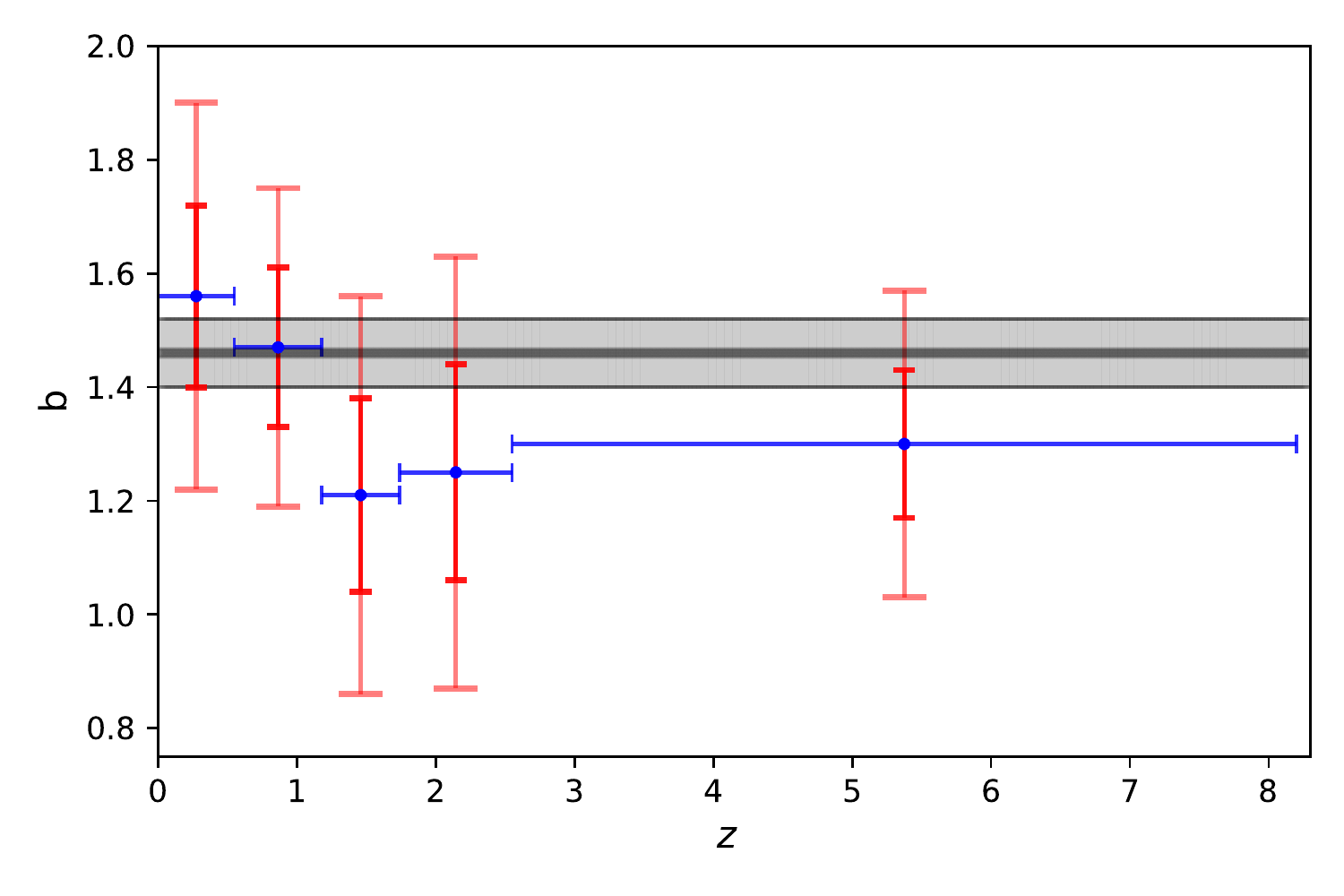}
	  \end{subfigure}

	\caption{The best-fit values (blue point), 1$\sigma$ uncertainties (solid red line) and 2$\sigma$ uncertainties (light red line) of coefficients $a$ (upper panel) and $b$ (lower panel) for each redshift bin. The black line and band are the best-fit values and 1 $\sigma$ uncertainties for the whole sample. }
	\label{Fabevo}       
    \end{figure}

	Whether the $E_{\mathrm {iso }}-E_{\mathrm{p}}$ correlation evolves with redshifts is important. Here we divide the full GRB sample into five redshift bins: [0-0.55], [0.55-1.18], [1.18-1.74], [1.74-2.55], [2.55-8.20]. The number of GRBs in each sub-sample are 20, 54, 44, 48 and 55, respectively. The best-fit values and 1$\sigma$ uncertainties of $E_{\mathrm {iso }}-E_{\mathrm{p}}$ correlation in each sub-sample are shown in Table \ref{T1}. Fig. \ref{Fabevo} shows the evolution of the coefficients at different redshift intervals. The results show that the values are in agreement with each other within 2$\sigma$ uncertainties, and $\sigma_{\mathrm{ext}}$ does not show an evolution trend in each bin.
	\par
	From Fig.\ref{Fabevo}, the best-fit values of $a$ go up and then down with the increase of redshifts, while the evolution of $b$ is opposite. Although there seems to be an evolutionary trend, they are consistent with each other at 2$\sigma$ level. Therefore, the $E_{\mathrm{iso }}-E_{\mathrm{p}}$ correlation is consistent for all redshift ranges. The correlation shows no significant evolution with redshifts, which is in line with \cite{2011MNRAS.415.3423W} and \cite{2021A&A...651L...8D}. If the correlation evolves with evolution, the method mentioned in \cite{2022MNRAS.514.1828D} can be used to fit the evolutionary function.
	\par

	\subsection{Calibrating the \EE{} correlation}\label{calibrating}
	During the calculation of $E_{\mathrm{iso }}$, the cosmological parameters are fixed as benchmark parameters \citep{2020A&A...641A...6P}. This may make the results depend on the choice of cosmological models \citep{2015NewAR..67....1W}. To avoid this circularity problem, we select some sub-samples of GRBs lying in a small redshift range. Among the GRBs in each sub-sample, the luminosity distances $d_{\mathrm{L}}$ are approximately same, that is why it can overcome the effect of cosmological models. Our selection criteria are:\par
	(1) The numbers of GRBs in each sub-sample should be large enough. A larger sample size would increase the reliability of the results and avoid selection bias whenever possible. \par
	(2) The extrinsic scatter of the fitting results should be small, because it indicates the quality of the fitting degree. So we prefer to select the sub-sample with relatively small $\sigma_{\mathrm{ext}}$, which also means that the $E_{\mathrm {iso }}-E_{\mathrm{p}}$ correlation in these groups of samples are better standardized. \par
	(3) The even distribution makes for a better fitting result. Points of each sub-samples are distributed on the $E_{\mathrm {iso }}-E_{\mathrm{p}}$ plane. We prefer to select points that are distributed evenly on the plane rather than concentrated on a small numerical range.\par
\begin{figure}
	\centering
	\includegraphics[width=0.5\textwidth,angle=0]{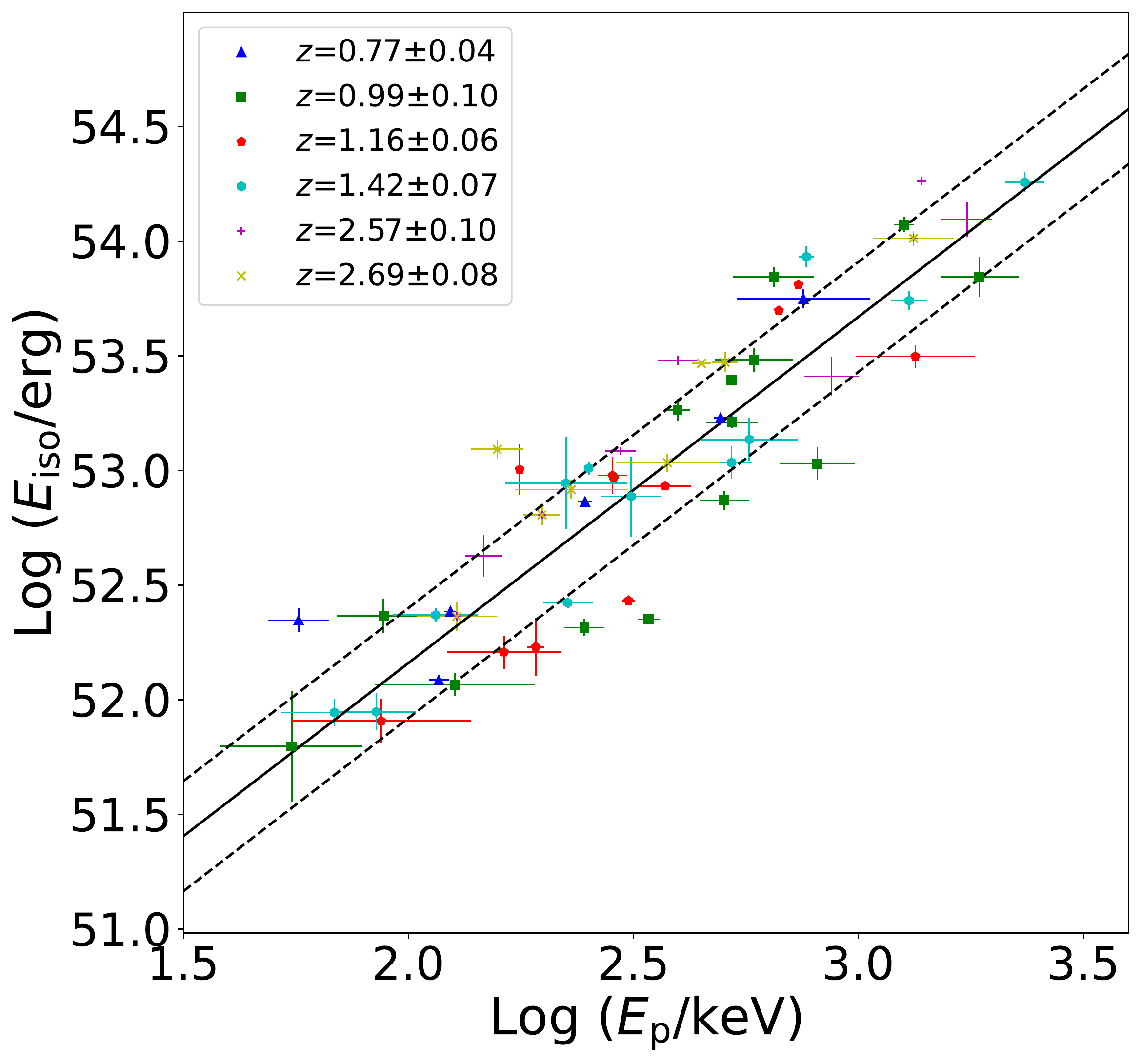}
	\caption{The fitting results in six sub-samples. The calibration result is shown as the solid black line with  1$\sigma$ uncertainty (the dotted black lines). }
	\label{Fdingbiao}       
    \end{figure}

	These six sub-samples are listed in Table \ref{T2}, and all have good fitting results. From Fig. \ref{Fdingbiao}, we can see that data from the fourth sub-sample distribute evenly on the $E_{\mathrm {iso }}-E_{\mathrm{p}}$ plane. In addition, the number of these data is relatively larger than other bins except for the second sub-sample. The extrinsic scatter of the fourth sub-sample is small. Therefore, we choose the fitting results of the fourth sub-sample: $a=49.14\pm0.45$, $b=1.51\pm0.17$ and $\sigma_{\mathrm{ext}}=0.24\pm0.08$. \cite{2019ApJ...873...39W} use the mock gravitational wave events associated with GRBs to get strict constraints on the parameters as  $a=52.93\pm0.04$, $b=1.41\pm0.07$ and $\sigma_{\mathrm{ext}}=0.39\pm0.03$.

    \subsection{The comparison with the methodology adopted in Dainotti fundamental plane relation}
    The Dainotti fundamental plane is the correlation among the peak prompt luminosity $L_{\mathrm{peak}}$, the X-ray luminosity of plateaus $L_X$, and the time at the end of the plateau emission $T^{*}_X(s)$ \citep{2016ApJ...825L..20D,2017A&A...600A..98D,2020ApJ...904...97D,2021PASJ...73..970D}, which is usually expressed as 
    \begin{equation}
    \log L_{X}=C_{o}+a \log T_{X}^{*}+b \log L_{\text {peak }}.
    \end{equation}
    This correlation is tight and can be used to constrain cosmological parameters \citep{2022PASJ..tmp...83D,2022MNRAS.514.1828D}. In the study of Dainotti fundamental plane, the screening criteria of long GRB sample are even more demanding \citep{2022MNRAS.tmp.2047C,2022ApJ...924...97W}. Therefore, the number in the sample is small. The $E_{\mathrm {iso }}-E_{\mathrm{p}}$ correlation focuses on the prompt emission, while the Dainotti correlation contains the characteristics of the afterglow emission. \cite{2022MNRAS.514.1828D} proved that GRB can be seen as cosmological distance indicators by analyzing the 3D Dainotti correlation based on the optical and X-ray sample. The determination on $\Omega_m$ from the optical sample is as efficacious as the X-ray one, making the optical plateau usable for cosmological applications.
    \par
    The selection bias and redshift evolution in GRB data may skew the analysis. In the process of fitting the Dainotti correlation, they use the Efron-Petrosian method to remove the evolution and recover the intrinsic relationships \citep{2021Galax...9...95D,2022MNRAS.514.1828D}. In this paper, we use the binning method to search the evolution of $E_{\mathrm {iso }}-E_{\mathrm{p}}$ correlation with redshifts. The consistency in five redshift bins shows no significant evolution for the correlation.  
    \par

	\section{Constraining cosmological models}\label{Sec4}
	\subsection{Cosmological models}
	To analyse the information of high-redshift universe carried by the GRB data, we consider $\Lambda$CDM and $w$CDM models. 
	For the $\Lambda$CDM model, the Hubble parameter is
	\begin{equation}
    H(z)=H_{0} \sqrt{\Omega_{m}(1+z)^{3}+\Omega_{k}(1+z)^{2}+\Omega_{\mathrm{\Lambda}}}.
    \end{equation}
    Since the constraint $\Omega_m + \Omega_k + \Omega_{\Lambda} = 1 $, $\Omega_{m} ,\Omega_{\Lambda}$ and $H_{0}$ are free parameters to be constrained in the $\Lambda$CDM model.
    \par
    In the $w$CDM model, the Hubble parameter is 
    \begin{equation}
    H(z)=H_{0} \sqrt{\Omega_{m}(1+z)^{3}+\Omega_{k}(1+z)^{2}+\Omega_{\mathrm{DE}}(1+z)^{3\left(1+w\right)}},
    \end{equation}
	where $\Omega_{\mathrm{DE}}$ is the dark energy density parameter and $w$ is the dark energy EOS parameter. In this parametrization, $w$ is a constant but $w \neq -1$. For the $w$CDM model, the free parameters are $\Omega_{m}, \Omega_{\mathrm{DE}}, w$ and $H_0$.
	\par
		\begin{figure*}
		\centering
		\subfigure[$H_0=67.4$ km s$^{-1}$ Mpc$^{-1}$]{
			\includegraphics[width=0.47\textwidth,angle=0]{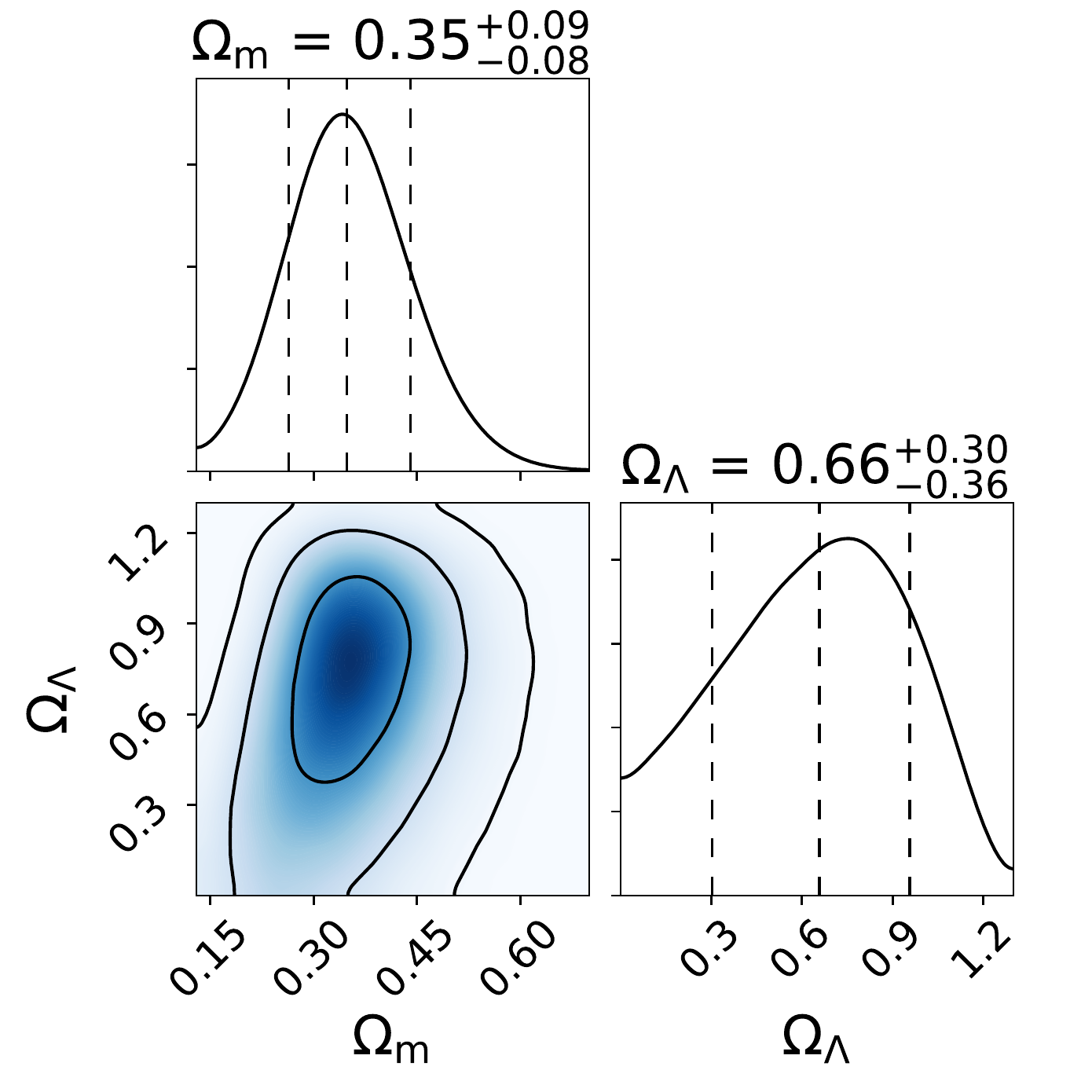}}
		\hfill
		\centering
		\subfigure[$H_0=73.2$ km s$^{-1}$ Mpc$^{-1}$]{
			\includegraphics[width=0.47\textwidth,angle=0]{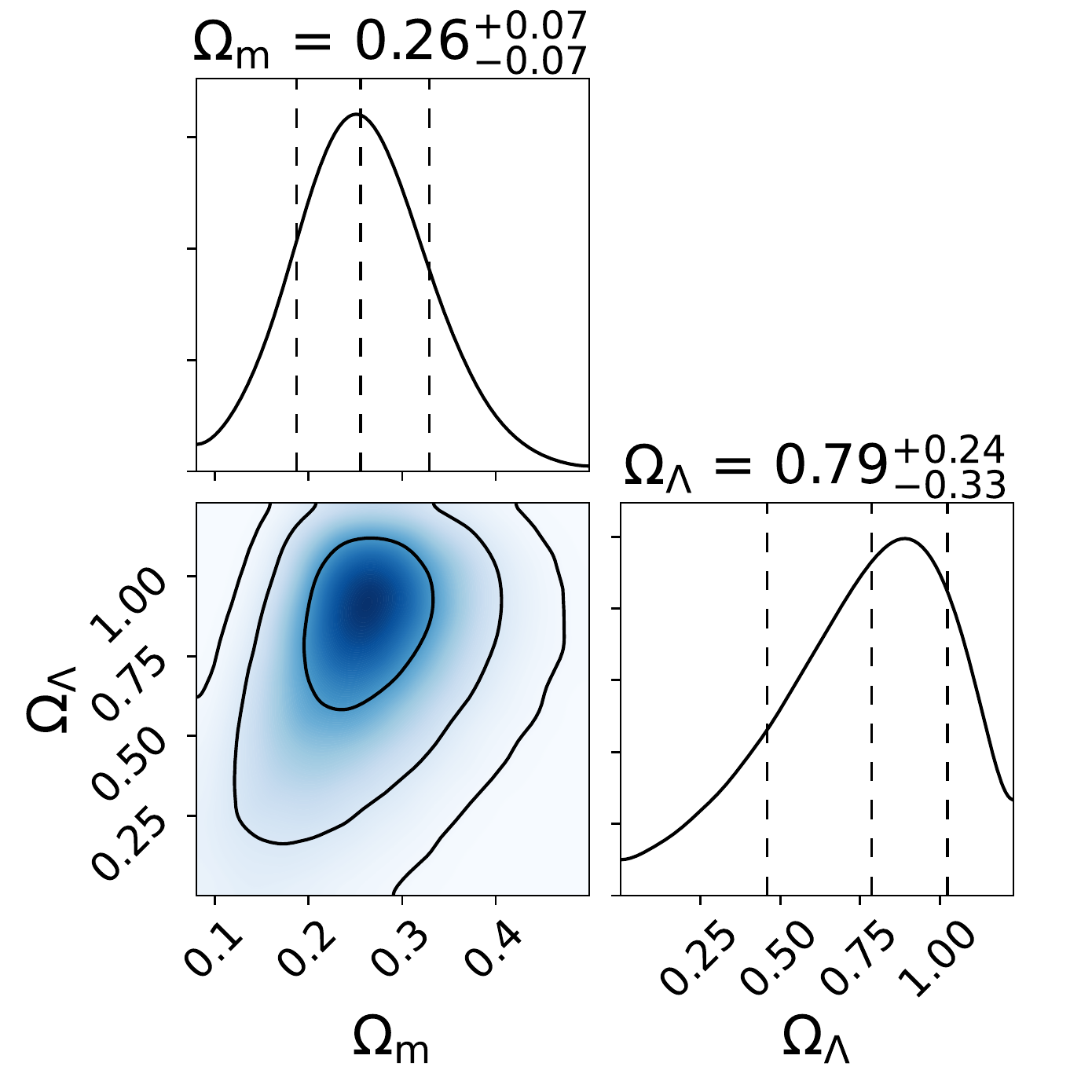}} 
		\caption{ The confidence regions from the GRB sample for different values of $H_0$. For the non-flat $\Lambda$CDM model, the confidence regions of the free parameters $\Omega_{m}$ and $\Omega_{\Lambda}$ are 1 $\sigma$, 2 $\sigma$ and 3$\sigma$ from the inner to the outer. }
		\label{FGRBnonflatLCDM}       
	\end{figure*}
	\subsection{Constraining cosmological models}
	The distance moduli of GRBs is calculated from $\mu=25+5 \log \left(d_{\mathrm{L}}/ \mathrm{Mpc}\right)$. So the distance moduli is
	\begin{equation}\label{distance moduli}
    \mu_{\mathrm{GRB}}=25+\frac{5}{2}\left[a+b \log E_{\mathrm{p}}-\log \frac{4 \pi S_{\mathrm{bolo}}}{(1+z)}\right].
    \end{equation}
    The propagated uncertainties of $E_{\mathrm{iso }}$ is given by the following equation
    \begin{equation}
    \sigma_{\log E_{\mathrm{iso }}}^{2}=\sigma_{a}^{2}+\left(\sigma_{b} \log \frac{E_{\mathrm{p}}}{\mathrm{keV}}\right)^{2}+\left(\frac{b}{\ln 10} \frac{\sigma_{E_{\mathrm{p}}}}{E_{\mathrm{p}}}\right)^{2}+\sigma_{\mathrm{ext }}^{2}.
    \end{equation}
    Then the propagated uncertainties of the distance moduli is calculated as
    \begin{equation}
    \sigma_{\mu}=\left[\left(\frac{5}{2} \sigma_{\log E_{\mathrm{iso }}}\right)^{2}+\left(\frac{5}{2 \ln 10} \frac{\sigma_{S_{\mathrm{bolo }}}}{S_{\mathrm {bolo }}}\right)^{2}\right]^{1 / 2}.
    \end{equation}
    These GRBs reveal the information of high-redshift universe, and their distance moduli are able to constrain cosmological models. Here we use the $\chi^{2}$ method to constrain the cosmological models mentioned above, and $\chi^{2}$ is 
    \begin{equation}
    \chi_{\mathrm{GRB}}^{2}=\sum_{i=1}^{N} \frac{\left[\mu_{\mathrm{GRB}}\left(z_{i}\right)-\mu\left(z_{i}\right)\right]^{2}}{\sigma^{2}_\mu (z_{i})},
    \end{equation}
    where $N$ is the number of the GRB sample, and $\mu_{\mathrm{GRB}}$ is the distance moduli calculated by Eq. (\ref{distance moduli}). For the MCMC analysis, the priors used for parameters are as follows: $\Omega_{m} \in$ [0,1], $H_0 \in$ [50,80], $\Omega_{\Lambda} \in$ [0,2] and $w \in$ [-5,0.33]. The GRB sample constrain cosmological parameters effectively. In order to get better limits, we also combine the GRB sample with the Pantheon SNe Ia sample \citep{2018ApJ...859..101S} to constrain cosmological models.
    \par
\begin{figure*}
	\centering
	  \subfigure[GRB + SNe Ia sample in flat $\Lambda$CDM]{
	  \includegraphics[width=0.45\textwidth,angle=0]{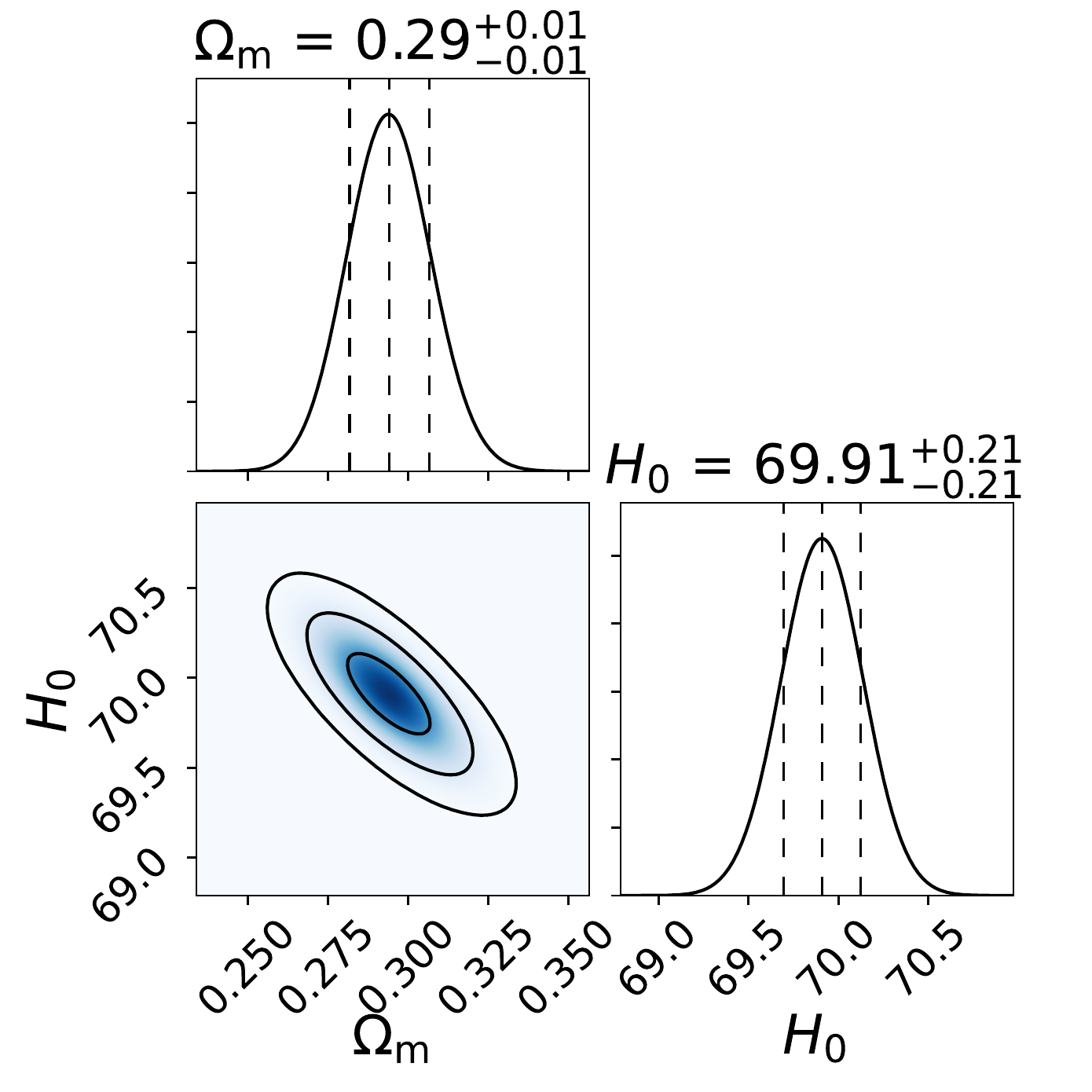}}
	  \hfill
	  \centering
	  \subfigure[GRB + SNe Ia sample in non-flat $\Lambda$CDM]{
	  \includegraphics[width=0.45\textwidth,angle=0]{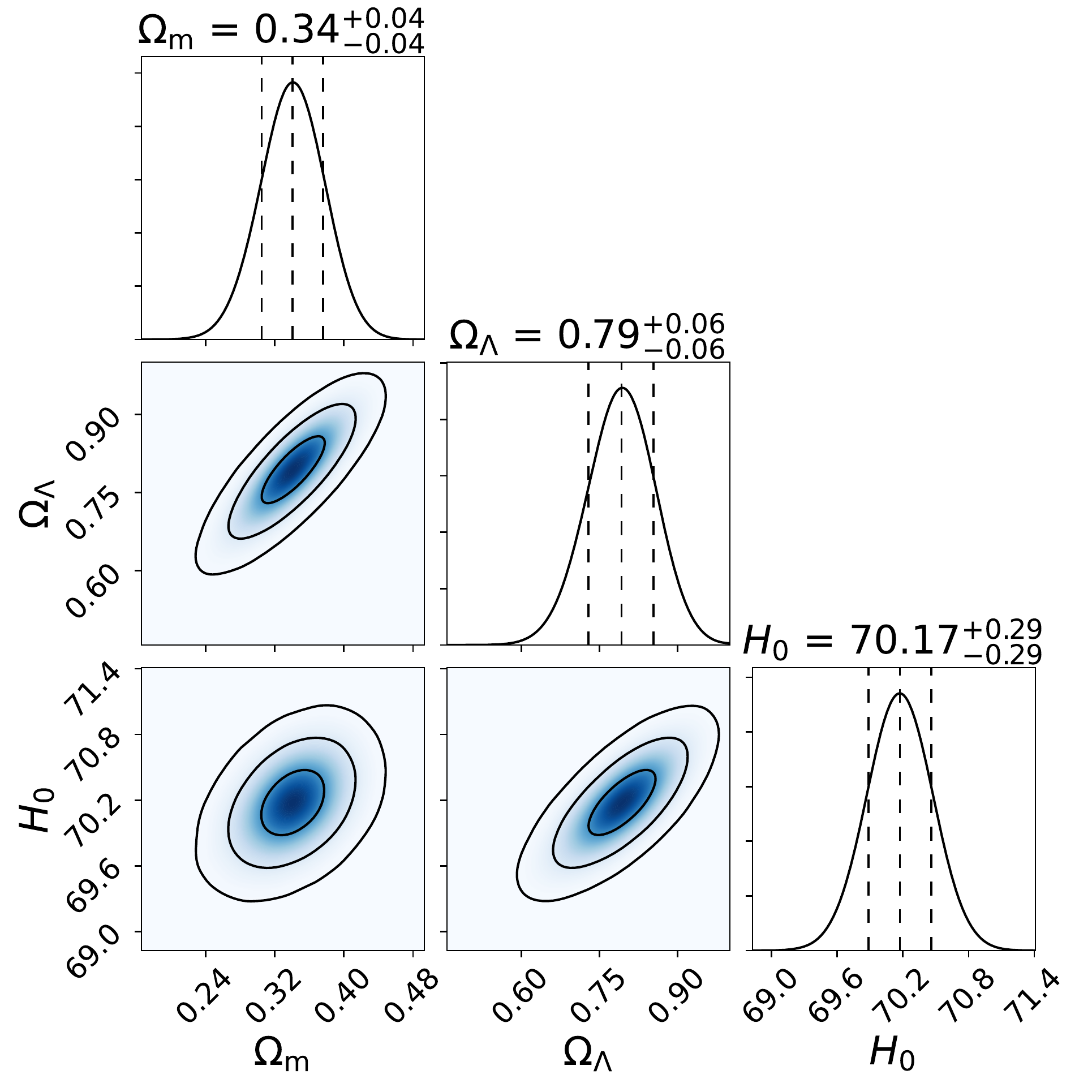}}
	\caption{ The confidence regions for the joint GRB and SNe Ia sample. The left and right panels are the results in flat and non-flat $\Lambda$CDM model, respectively.}
	\label{FGRBSN_LCDM}       
    \end{figure*}

    For the non-flat $\Lambda$CDM model, the Hubble constant $H_0$ is first fixed as 67.4 km s$^{-1}$ Mpc$^{-1}$ \citep{2020A&A...641A...6P} and then 73.2 km s$^{-1}$ Mpc$^{-1}$ \citep{2021ApJ...908L...6R}. The best-fit results are $\Omega_{m}=0.35^{+0.09}_{-0.08}$ and $\Omega_{\Lambda}=0.66^{+0.30}_{-0.36}$ with 1$\sigma$ uncertainties when $H_0=67.4$ km s$^{-1}$ Mpc$^{-1}$, $\Omega_{m}=0.26\pm0.07$ and $\Omega_{\Lambda}=0.79^{+0.24}_{-0.33}$ when $H_0=73.2$ km s$^{-1}$ Mpc$^{-1}$. The fitting results of the GRB sample are shown in Fig. \ref{FGRBnonflatLCDM}. The GRB sample is combined with the Pantheon sample to get better limits, the results of which are $\Omega_{m}=0.34\pm0.04$, $\Omega_{\Lambda}$=$0.79\pm0.06$ and $H_0=70.17\pm0.29$ km s$^{-1}$ Mpc$^{-1}$. For the flat $\Lambda$CDM model, the results obtained by combining GRB data with SNe Ia data are better than those obtained by GRB data alone. The best-fit results are $\Omega_{m}=0.29\pm0.01$ and $H_0=69.91\pm0.21$ km s$^{-1}$ Mpc$^{-1}$ for the joint data. The results of the joint data are shown in Fig. \ref{FGRBSN_LCDM}. In addition, the value of $\Omega_{m}$ is consistent with the constraints from SNe Ia \citep{2018ApJ...859..101S} and CMB \citep{2020A&A...641A...6P} within 1$\sigma$ range. In \cite{2022MNRAS.510.2928C}, they fit the parameters of $E_{\mathrm {iso }}-E_{\mathrm{p}}$ correlation and $\Omega_m$ simultaneously. The result is $\Omega_m>0.247$ in flat $\Lambda$CDM model, $\Omega_m>0.287$ and $\Omega_k=0.694^{+0.626}_{-0.848}$ for the non-flat $\Lambda$CDM model. The lower limits on the matter density parameter are consistent with currently accelerating cosmological expansion. The three-parameter fundamental plane relation in \cite{2022MNRAS.512..439C} provided $\Omega_m>0.411$ in flat $\Lambda$CDM model and $\Omega_m>0.491$ for the non-flat $\Lambda$CDM model. After being combined with OHD, BAO and GRBs, $\Omega_m=0.300^{+0.016}_{-0.018}$ and $\Omega_m=0.293\pm0.023$ were found in the flat and non-flat $\Lambda$CDM models, respectively.

    \par
\begin{figure*}
	\centering
	  \subfigure[GRB + SNe Ia sample in flat $w$CDM ]{
	  \includegraphics[width=0.45\textwidth,angle=0]{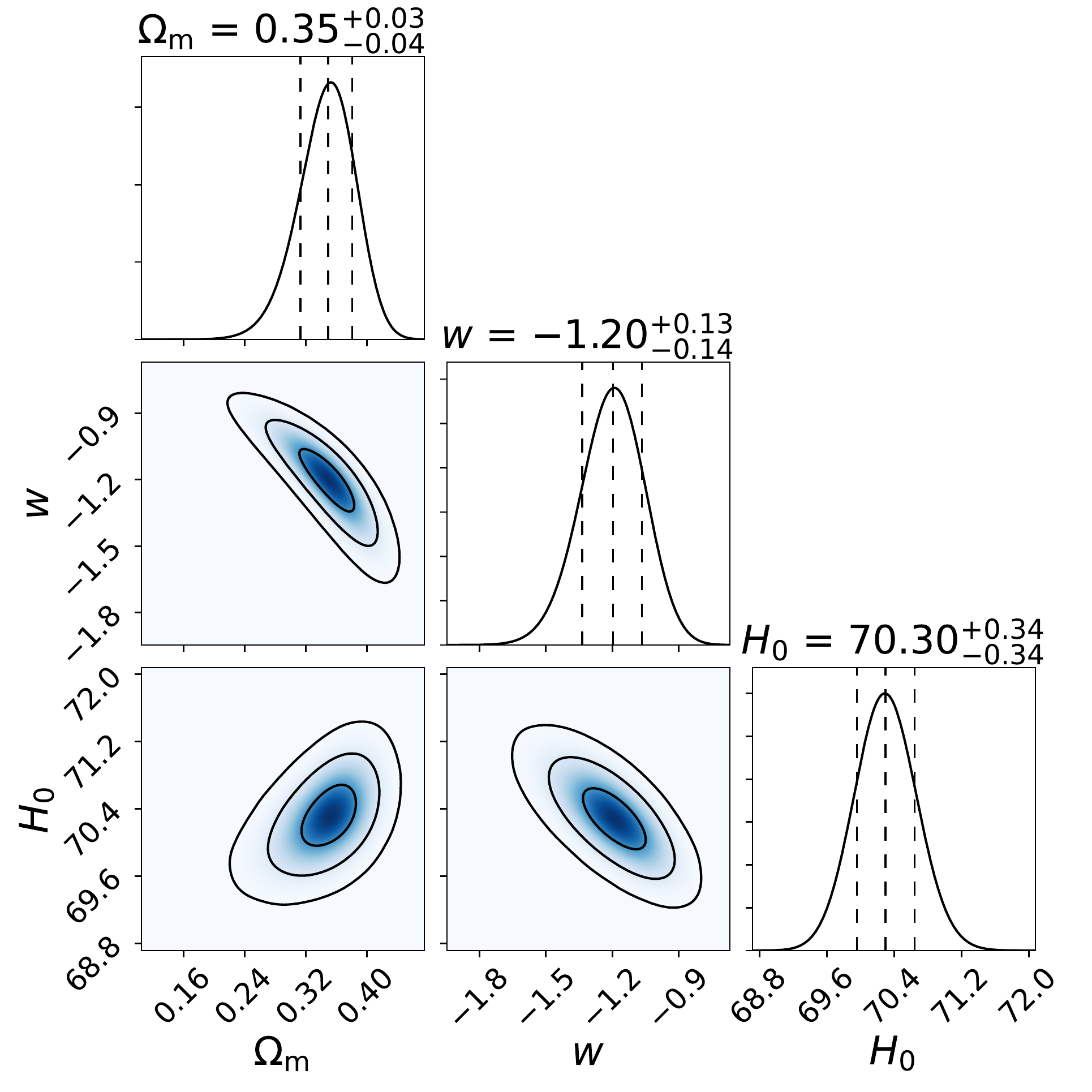}}
	  \hfill
	  \centering
	  \subfigure[GRB + SNe Ia sample in non-flat $w$CDM]{
	  \includegraphics[width=0.45\textwidth,angle=0]{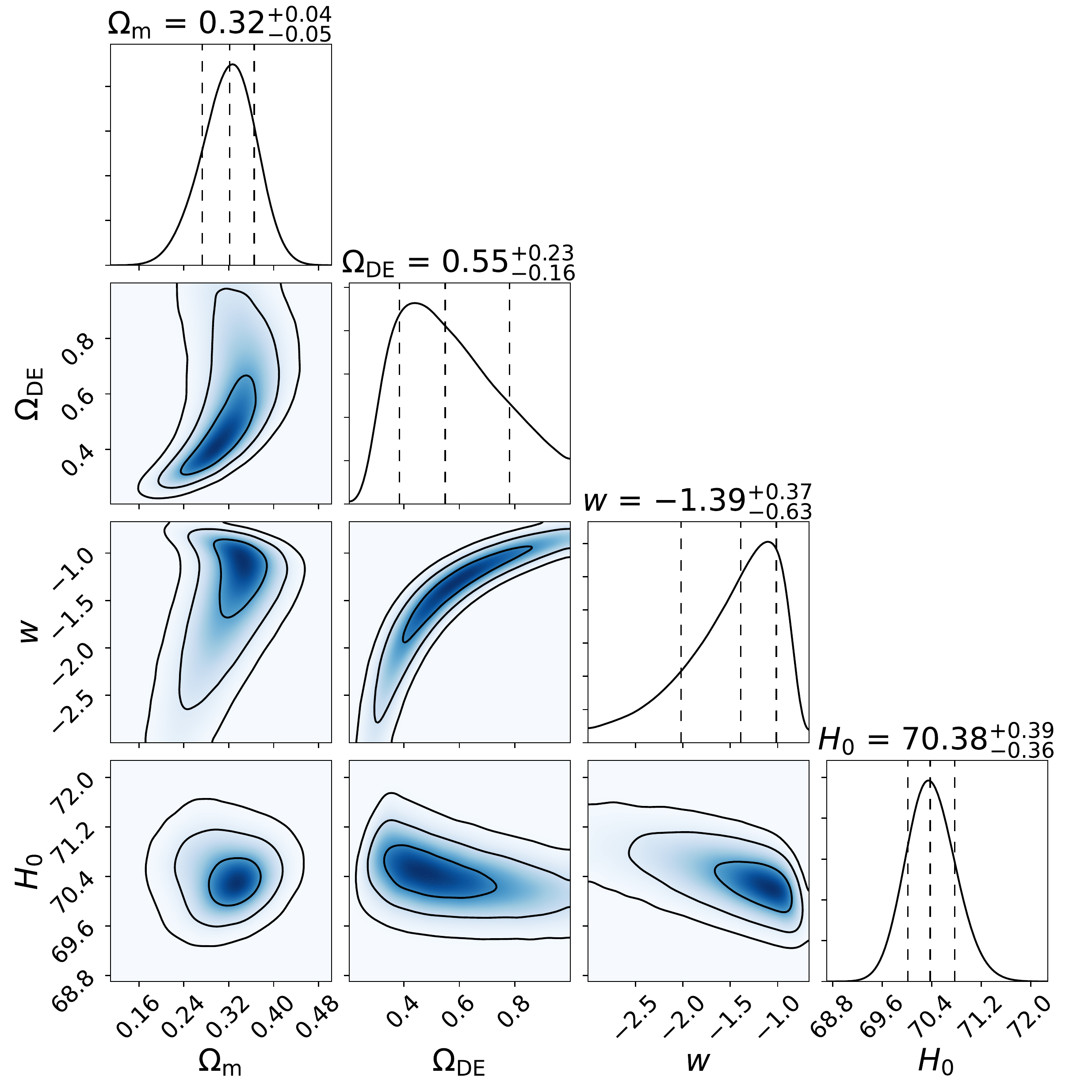}}
	\caption{ The confidence regions for the joint GRB and SNe Ia sample. The left and right panels are the results in flat and non-flat $w$CDM model, respectively. }
	\label{FwCDM}       
    \end{figure*}

    For the non-flat $w$CDM model, the sample combined with GRB data and SNe Ia data constrain the cosmological parameters as $\Omega_{m}=0.32^{+0.04}_{-0.05}$, $\Omega_{\mathrm{DE}}=0.55^{+0.23}_{-0.16}$, $w=-1.39^{+0.37}_{-0.63}$ and $H_0=70.32^{+0.39}_{-0.36}$ km s$^{-1}$ Mpc$^{-1}$. For the flat $w$CDM, the results are $\Omega_{m}=0.35^{+0.03}_{-0.04}$, $w=-1.20^{+0.13}_{-0.14}$ and $H_0=70.30\pm0.34$ km s$^{-1}$ Mpc$^{-1}$. The fitting results of the joint data are shown in Fig. \ref{FwCDM}. \cite{2022MNRAS.512..439C} provide $\Omega_m=0.282^{+0.023}_{-0.021}$, $w=-0.731^{+0.150}_{-0.096}$ and $H_0= 65.54^{+2.26}_{-2.58}$ km s$^{-1}$ Mpc$^{-1}$ in the flat $w$CDM model. The constraints from OHD and BAO trend to a low value of $H_0$.

   \par
   \section{The evolution of dark energy EOS}\label{Sec5}
	To study the evolution of dark energy EOS, a flat universe with an evolving dark energy EOS is considered. According to the observations from Planck \citep{2020A&A...641A...6P}, the assumption of flatness is reasonable. The EOS of dark energy is $w = p/\rho$, where $p$ is the pressure and $\rho$ is the energy density. The EOS $w$ is a remarkable characterization of dark energy. For revealing dark energy, it is crucial to research whether and how it evolves over time. In order to avoid adding some priors on the nature of dark energy, a non-parametric approach is used here \citep{2003PhRvL..90c1301H,2005PhRvD..71b3506H}.
	\par
	From the Friedmann equation, the expansion rate in a flat universe is expressed as
	\begin{equation}\label{Hubble expansion rate}
    \frac{H^{2}(z)}{H_{0}^{2}}=\Omega_{m}(1+z)^{3}+\Omega_{\mathrm{DE}}f(z) ,
    \end{equation}
    where $f(z) = \exp \left(3 \int \frac{d z^{\prime}}{1+z^{\prime}}\left[1+w\left(z^{\prime}\right)\right]\right)$, $\Omega_{\mathrm{DE}} = 1-\Omega_{m}$ is the dark energy density parameter at present and $w$ is the parameter, which describes the properties of dark energy EOS. The function $f(z)$ is related to the evolution of dark energy EOS at different redshifts. If we split the function up into several redshift bins and consider $w(z)$ is a constant in each redshift bin, then $f(z)$ becomes a piece-wise function and is described as 
	\begin{equation}
    f\left(z_{n-1}<z \leq z_{n}\right)=(1+z)^{3\left(1+w_{n}\right)} \prod_{i=0}^{n-1}\left(1+z_{i}\right)^{3\left(w_{i}-w_{i+1}\right)}.
    \end{equation}
	The parameter $w_i$ is the EOS $w(z)$ in the $i$th redshift bin, $n$ is the serial number of the redshift bin, and the zeroth bin is defined as $z_0$ = 0. Here we add an assumption that the EOS is fixed as $w = -1$ at $z > 8.2$ without affecting the fitting results \citep{2014PhRvD..89b3004W}.
	\par
	In this parameterization, no assumptions are made about the nature of dark energy, since different parameters are introduced in each redshift bin. When choosing the number and range of redshift bins, the limitation from the whole sample should be taken into account. Redshift intervals for each bin are determined in the process of separating redshift bins. First, we find that the number of data in each bin should be big enough to get a strict constraint on EOS $w_i$. This implies that in order to avoid poor restrictions, we have to choose loose intervals as the number of GRBs decreases with increasing redshifts. Second, the magnitude of each redshift interval should be reasonable. A too loose redshift interval may conflict with the approximation that $w(z)$ is a constant in each redshift bin. Finally, we expect the amount of data in each bin to be as equal as possible. After testing many kinds of redshift bins, we finally choose 11 bins in this analysis. The upper boundaries are $z_i$ = 0.1, 0.2, 0.3, 0.4, 0.5, 0.6, 0.7, 0.8, 0.9, 1.3, 8.2. We have to adopt a large redshift interval for the last redshift bin due to the lack of data at high redshifts.
	\par
	The MCMC method mentioned above is used to fit $w_i$ in each redshift bin. Due to the function $f(z)$ depends on the summation of $w_i$ over redshift, the EOS parameters $w_i$ are correlated. For the sake of removing the correlation, the covariance matrix of $w_i$ is calculated as
	\begin{equation}
    \mathbf{C}=\langle \mathbf{w} \textbf{w}^{\rm T}\rangle-\langle\mathbf{w}\rangle\langle\mathbf{w}^{\rm T}\rangle,
    \end{equation}
	where $\mathbf{w}$ is a vector with components $w_i$. It is not diagonal, but through multiplying a transformation matrix, we obtain a set of decorrelated parameters
	\begin{equation}
	\centering
	\tilde{\mathbf{w}}=\mathbf{Tw},
	\end{equation}
	in which $\tilde{\mathbf{w}}$ is the uncorrelated dark energy parameters with components $\tilde{{w_i}}$. The transformation can be computed as \cite{2005PhRvD..71b3506H}. First the Fisher matrix is
	\begin{equation}
    \mathbf{F} \equiv \mathbf{C}^{-1} \equiv \mathbf{O}^{\mathrm{T}} \mathbf{\Lambda} \mathbf{O} ,
    \end{equation}
	where $\mathbf{\Lambda}$ is diagonal. Then the transformation matrix $\mathbf{T}$ is defined as 
	\begin{equation}
    \mathbf{T}=\mathbf{O}^{\mathrm{T}} \mathbf{\Lambda}^{\frac{1}{2}} \mathbf{O} .
    \end{equation}
	The transformation $\mathbf{T}$ is normalized so that its rows, which represents the weights for $w_i$, sum to unity. Another advantage of this transformation is that the weights are almost positive everywhere .
	\par
	The method mentioned above is used in conjunction with a joint data set of the latest observations including the GRB sample, CMB from Planck, SNe Ia, and the OHD. For SNe Ia data, the Pantheon sample from \cite{2018ApJ...859..101S} are used. The distance priors are taken from \cite{2019JCAP...02..028C}, such as CMB shift parameters $R=1.7502\pm 0.0046$, $l_A=301.471^{+0.089}_{-0.090}$ and $\Omega_bh^2=0.02236\pm0.00015$. The definitions of the distance priors are as follows
    \begin{equation}
    R\left(z_{*}\right) \equiv \frac{\left(1+z_{*}\right) D_{\mathrm{A}}\left(z_{*}\right) \sqrt{\Omega_{m} H_{0}^{2}}}{c},
    \end{equation}
    \begin{equation}
    l_{\mathrm{A}}=\left(1+z_{*}\right) \frac{\pi D_{\mathrm{A}}\left(z_{*}\right)}{r_{s}\left(z_{*}\right)},
    \end{equation}
	in which $z_*$ is the redshift at the photon decoupling epoch, $D_{A}$ is the angular diameter distance, and $r_s$ is the comoving sound horizon. For the OHD, the data from \cite{2018ApJ...856....3Y} are adopted.
	\par
\begin{figure}
	\centering
	\includegraphics[width=0.5\textwidth,angle=0]{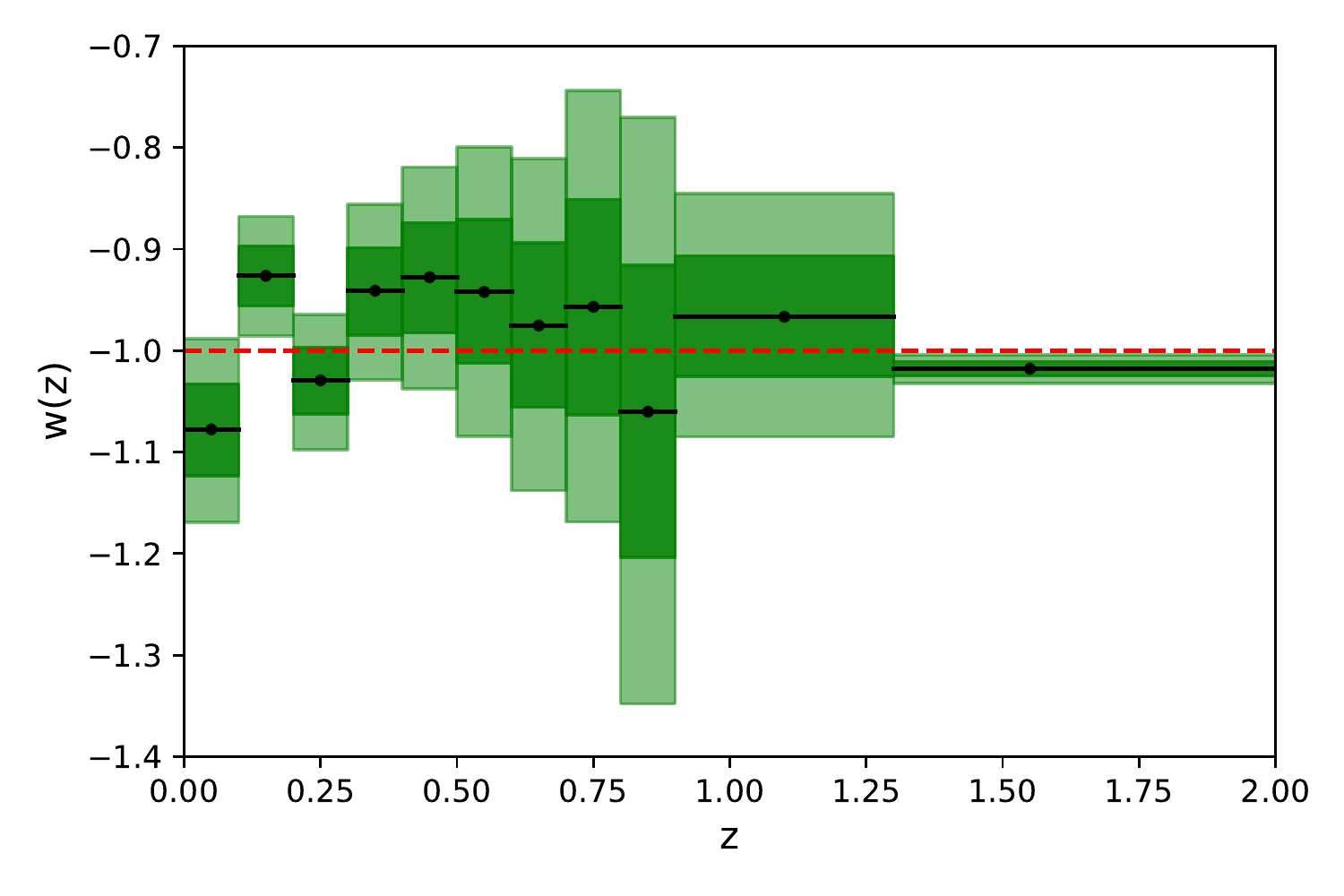}
	\caption{The evolution of EOS $w_i$ at different redshift bins. The red dotted line shows $w = -1$. The green band is the uncertainty (1$\sigma$ and 2$\sigma$ from the darker to the lighter) of the best-fit value (black point). }
	\label{Fwevo}       
    \end{figure}

	During the fitting process, $\Omega_{m}$ and $H_0$ are taken as free parameters, so there are 13 cosmological parameters to be constrained. The final results are $\Omega_{\mathrm{m}}=0.26\pm0.01$, $H_0=70.64^{+0.39}_{-0.38}$ km s$^{-1}$ Mpc$^{-1}$. And the uncorrelated dark energy EOS parameters $w_i$ at different redshift bins are shown in Fig.\ref{Fwevo}. In \cite{2022PASJ..tmp...83D}, they provided $\Omega_m=0.321\pm0.003$ and $H_0=69.644\pm0.116$ km s$^{-1}$ Mpc$^{-1}$ for a BAOs+SNe Ia+GRBs sample. We also find that the joint sample can improve the precision of the result. The effect of the number in each bin is taken into account in determining the redshift interval.
	\par
	Combined with the cosmological models mentioned above, the results are used to check whether the $\Lambda$CDM model is still the best candidate. The dark energy EOS is equal to -1 for $\Lambda$CDM model but a function of redshifts in dynamical dark energy models. The results show an evolutionary trend to deviate from the $\Lambda$CDM model. But within 2$\sigma$ uncertainties, the results of EOS $w_i$ are still consistent with -1 except for the second bin. The dark energy EOS evolves with redshifts and crosses the -1 boundary similar to previous investigations \citep{2014PhRvD..89b3004W,2017NatAs...1..627Z}. 
	\par
	It is worth noting that the dark energy EOS seems to be oscillating among the first four bins. What is more, it crosses the -1 boundary with the increase of redshifts, which is not permitted in the $\Lambda$CDM model. This may be a clue to the dynamical dark energy models, although most observations support the $\Lambda$CDM model. The data seem to prefer an upward tendency at redshifts $0.2 < z < 0.5$, which is consistent with \cite{2014PhRvD..89b3004W}. And the best-fit values of dark energy EOS parameters $w_i$ are all greater than -1 at $0.3 < z < 0.8$, showing no difference with \cite{2009MNRAS.398L..78Q}. For the last bin, the error is very small, although the redshift interval is from 1.3 to 8.2. It may be due to the fact that this bin contains more OHD data than others. In order to reduce the range of the last bin, more high redshift observational data are needed.
	 \par
	 We also notice that the errors will be smaller if we fix the cosmological parameters $\Omega_m$ and $H_0$. But this will add some priors on the cosmological model and significantly affect the final fitting results. Considering the Hubble tension between the value of $H_0$ from Cepheids ($H_0=73.2\pm1.3$ km s$^{-1}$ Mpc$^{-1}$; \citealt{2021ApJ...908L...6R}) and the value of $H_0$ from CMB ($H_0=67.4\pm0.5$ km s$^{-1}$ Mpc$^{-1}$; \citealt{2020A&A...641A...6P}), it is difficult to determine a specific value of $H_0$ and we final decide to free it. Furthermore, our fitting results of $H_0$ are consistent with \cite{2021ApJ...908L...6R} within 2$\sigma$ ranges. This may because the main data of our analysis come from the local observations.
	 \par

	\section{Conclusions and Discussion  }\label{Sec6}
	In this paper, a sample including 221 long GRBs is compiled for the $E_{\mathrm{iso}}-E_{\mathrm{p}}$ correlation. Fitting this correlation with the sample, we obtain $a=49.24\pm0.16$, $b=1.46\pm0.06$ and $\sigma_{\mathrm{ext}}=0.39\pm0.02$. Then, the possible redshift evolution of $E_{\mathrm{iso}}-E_{\mathrm{p}}$ correlation is studied in five redshift bins. The results show that the correlation does not show significant evolution with redshifts in 2$\sigma$ uncertainties. The correlation is calibrated by GRBs in a small redshift range, which is model-independent. The calibrated results are $a=49.14\pm0.45$, $b=1.51 \pm0.17$ and $\sigma_{\mathrm{ext}}=0.24\pm0.08$. The parameters are consistent with the results fitted by the whole GRB sample within $1\sigma$ confidence level, which may also confirms that the correlation does not evolve with redshifts.
	\par
	With the calibrated correlation, the sample is used to constrain cosmological parameters. Here, we consider $\Lambda$CDM and $w$CDM cosmological models. In order to get better constraints, the sample is combined with SNe Ia data. The results show that the combination of GRBs data and SNe Ia data constrain the cosmological parameters better. The fitting results support the $\Lambda$CDM model. 
	\par
	In order to study the physical properties of dark energy, we use \textbf{a} non-parametric approach. Eleven redshift bins are used in this work due to the abundance of data. Our result shows that there is a hint for dynamical energy models. The evolution of dark energy EOS $w_i$ has a tendency to deviate from $-1$. It is oscillating at low redshift and consistent with the $\Lambda$CDM model at high redshift at 2$\sigma$ confidence level.  Compared with previous works, the GRBs data fills the gap between SNe Ia and CMB. There are more than half of GRBs at redshift $z > 1.5$, helping to constrain the EOS more strictly. The deviation from $-1$ in some bins is a weak hint for the dynamical dark energy models. 
	\par
	In the future, as more GRBs will be detected, some correlations will be found and current correlation can be improved. We are looking forward to the observations by the French-Chinese satellite space-based multi-band astronomical variable objects monitor (SVOM) \citep{2016arXiv161006892W}, the Einstein Probe (EP) \citep{2015arXiv150607735Y} and the Transient High-Energy Sky and Early Universe Surveyor (THESEUS) \citep{2018AdSpR..62..191A} to help us explore high-redshift universe using GRBs, such as cosmic expansion, reionization and metal enrichment history \citep{Wang2012}. 

\section*{Acknowledgements}
This work was supported by the National Natural Science
Foundation of China (grant No. U1831207), the China Manned Spaced Project (CMS-CSST-2021-A12), Jiangsu Funding Program for Excellent Postdoctoral Talent (20220ZB59).

\section*{Data Availability}

The data that support the findings of this study are available in Table \ref{GRBsample}.



\bibliographystyle{mnras}
\bibliography{MNRAS} 



\newpage
\onecolumn
{\small
\begin{longtable}{@{} c @{ } c @{ } c @{ } c @{ }  c }
 \hline
 GRB    &   Redshift  &  $E_{\rm p}$(keV)  &  $E_{\rm iso}^{\mathrm{(a)}}$($10^{52}$ erg)    &  Refs. $^{\mathrm{(b)}}$\\
\hline
\endhead
060218 & 0.034 & 4.90 $\pm$ 0.30 & 0.0054 $\pm$ 0.0003 & (1) \\  \hline
180728 & 0.117 & 87.04 $\pm$ 1.95 & 0.28 $\pm$ 0.001 & (5) \\  \hline 
060614 & 0.125 & 55.00 $\pm$ 45.00 & 0.22 $\pm$ 0.09 & (1) \\  \hline 
030329 & 0.17 & 100.00 $\pm$ 23.00 & 1.48 $\pm$ 0.26 & (1) \\  \hline 
020903 & 0.25 & 3.37 $\pm$ 1.79 & 0.0024 $\pm$ 0.0006 & (1) \\  \hline 
130427A & 0.34 & 1112.20 $\pm$ 6.70 & 95.10 $\pm$ 30.10 & (4) \\  \hline 
011121 & 0.36 & 1060.00 $\pm$ 275.00 & 7.97 $\pm$ 2.19 & (1) \\  \hline 
020819 & 0.41 & 70.00 $\pm$ 21.00 & 0.69 $\pm$ 0.18 & (1) \\  \hline 
101213 & 0.414 & 440.00 $\pm$ 180.00 & 2.72 $\pm$ 0.53 & (3) \\  \hline 
190114 & 0.424 & 1477.49 $\pm$ 17.31 & 36.87 $\pm$ 0.02 & (5) \\  \hline 
990712 & 0.434 & 93.00 $\pm$ 15.00 & 0.69 $\pm$ 0.13 & (1) \\  \hline 
010921 & 0.45 & 129.00 $\pm$ 26.00 & 0.97 $\pm$ 0.09 & (1) \\  \hline 
130831A & 0.48 & 81.35 $\pm$ 5.92 & 0.80 $\pm$ 0.30 & (4) \\  \hline 
091127 & 0.49 & 51.00 $\pm$ 5.00 & 1.65 $\pm$ 0.18 & (3) \\  \hline 
081007 & 0.53 & 61.00 $\pm$ 15.00 & 0.18 $\pm$ 0.02 & (2) \\  \hline 
090618 & 0.54 & 250.41 $\pm$ 4.47 & 28.59 $\pm$ 0.52 & (3) \\  \hline 
100621 & 0.54 & 146.49 $\pm$ 23.90 & 4.60 $\pm$ 2.00 & (4) \\  \hline 
060729 & 0.543 & 77.00 $\pm$ 38.00 & 0.42 $\pm$ 0.09 & (1) \\  \hline 
090424 & 0.544 & 249.97 $\pm$ 3.32 & 4.07 $\pm$ 0.35 & (2) \\  \hline 
101219 & 0.55 & 108.00 $\pm$ 12.00 & 0.63 $\pm$ 0.06 & (3) \\  \hline 
170607 & 0.557 & 174.06 $\pm$ 9.03 & 1.10 $\pm$ 0.03 & (5) \\  \hline 
130215 & 0.6 & 247.54 $\pm$ 100.61 & 4.70 $\pm$ 2.40 & (4) \\  \hline 
050525 & 0.606 & 129.00 $\pm$ 6.50 & 2.29 $\pm$ 0.49 & (1) \\  \hline 
110106 & 0.618 & 194.00 $\pm$ 56.00 & 0.73 $\pm$ 0.07 & (3) \\  \hline 
131231 & 0.642 & 292.42 $\pm$ 4.03 & 23.76 $\pm$ 0.33 & (5) \\  \hline 
161129 & 0.645 & 240.84 $\pm$ 42.61 & 1.84 $\pm$ 0.25 & (5) \\  \hline 
050416 & 0.653 & 22.00 $\pm$ 4.50 & 0.11 $\pm$ 0.018 & (1) \\  \hline 
180720 & 0.654 & 1052.01 $\pm$ 15.43 & 56.57 $\pm$ 1.05 & (5) \\  \hline 
111209 & 0.68 & 519.87 $\pm$ 88.88 & 87.70 $\pm$ 36.10 & (4) \\  \hline 
080916 & 0.689 & 208.00 $\pm$ 11.00 & 0.98 $\pm$ 0.09 & (2) \\  \hline 
020405 & 0.69 & 354.00 $\pm$ 10.00 & 10.64 $\pm$ 0.89 & (1) \\  \hline 
970228 & 0.695 & 195.00 $\pm$ 64.00 & 1.65 $\pm$ 0.12 & (1) \\  \hline 
991208 & 0.706 & 313.00 $\pm$ 31.00 & 22.97 $\pm$ 1.86 & (1) \\  \hline 
041006 & 0.716 & 98.00 $\pm$ 20.00 & 3.11 $\pm$ 0.89 & (1) \\  \hline 
140512 & 0.725 & 1191.99 $\pm$ 58.24 & 9.21 $\pm$ 4.64 & (5) \\  \hline 
090328 & 0.736 & 1157.91 $\pm$ 55.55 & 14.18 $\pm$ 0.99 & (2) \\  \hline 
160804 & 0.736 & 123.93 $\pm$ 4.18 & 2.43 $\pm$ 0.23 & (5) \\  \hline 
150821 & 0.755 & 493.55 $\pm$ 17.11 & 16.92 $\pm$ 0.83 & (5) \\  \hline 
030528 & 0.78 & 57.00 $\pm$ 9.00 & 2.22 $\pm$ 0.27 & (1) \\  \hline 
051022 & 0.8 & 754.00 $\pm$ 258.00 & 56.04 $\pm$ 5.34 & (1) \\  \hline 
100816 & 0.805 & 246.72 $\pm$ 8.48 & 7.30 $\pm$ 0.02 & (3) \\  \hline 
150514 & 0.807 & 116.74 $\pm$ 5.91 & 1.22 $\pm$ 0.08 & (5) \\  \hline 
151027 & 0.81 & 364.54 $\pm$ 24.47 & 5.16 $\pm$ 0.37 & (5) \\  \hline 
110715 & 0.82 & 218.40 $\pm$ 20.93 & 5.10 $\pm$ 1.60 & (4) \\  \hline 
970508 & 0.835 & 145.00 $\pm$ 43.00 & 0.61 $\pm$ 0.13 & (1) \\  \hline 
990705 & 0.842 & 459.00$\pm$ 139.00 & 18.70 $\pm$ 2.67 & (1) \\  \hline 
000210 & 0.846 & 753.00 $\pm$ 26.00 & 15.41 $\pm$ 1.69 & (1) \\  \hline 
040924 & 0.859 & 102.00 $\pm$ 35.00 & 0.98 $\pm$ 0.09 & (1) \\  \hline 
170903 & 0.886 & 179.29 $\pm$ 13.39 & 0.87 $\pm$ 0.91 & (5) \\  \hline 
140506 & 0.889 & 371.53 $\pm$ 25.30 & 1.10 $\pm$ 0.35 & (5) \\  \hline 
091003 & 0.897 & 810.00 $\pm$ 157.00 & 10.70 $\pm$ 1.78 & (3) \\  \hline 
141225 & 0.915 & 341.55 $\pm$ 19.28 & 2.24 $\pm$ 0.31 & (5) \\  \hline 
080319B & 0.937 & 1261.00 $\pm$ 65.00 & 117.87 $\pm$ 8.93 & (1) \\  \hline 
071010 & 0.947 & 88.00 $\pm$ 21.00 & 2.32 $\pm$ 0.40 & (1) \\  \hline 
970828 & 0.958 & 586.00 $\pm$ 117.00 & 30.38 $\pm$ 3.57 & (1) \\  \hline 
980703 & 0.966 & 503.00 $\pm$ 64.00 & 7.42 $\pm$ 0.71 & (1) \\  \hline 
091018 & 0.971 & 55.00 $\pm$ 20.00 & 0.63 $\pm$ 0.35 & (3) \\  \hline 
021211 & 1.01 & 127.00 $\pm$ 52.00 & 1.16 $\pm$ 0.13 & (1) \\  \hline 
991216 & 1.02 & 648.00 $\pm$ 134.00 & 69.79 $\pm$ 7.16 & (1) \\  \hline 
140508 & 1.027 & 521.76 $\pm$ 12.12 & 24.87 $\pm$ 0.87 & (5) \\  \hline 
080411 & 1.03 & 524.00 $\pm$ 70.00 & 16.19 $\pm$ 0.98 & (1) \\  \hline 
000911 & 1.06 & 1856.00 $\pm$ 371.00 & 69.86 $\pm$ 14.33 & (1) \\  \hline 
091208 & 1.063 & 246.00 $\pm$ 25.00 & 2.06 $\pm$ 0.18 & (3) \\  \hline 
091024 & 1.092 & 396.22 $\pm$ 25.31 & 18.38 $\pm$ 1.99 & (3) \\  \hline 
980613 & 1.096 & 194.00 $\pm$ 89.00 & 0.61 $\pm$ 0.09 & (1) \\  \hline 
080413B & 1.1 & 163.00 $\pm$ 47.50 & 1.61 $\pm$ 0.27 & (2) \\  \hline 
201216 & 1.1 & 735.40 $\pm$ 10.24 & 64.59 $\pm$ 0.02 & (5) \\  \hline 
981226 & 1.11 & 87.00 $\pm$ 40.00 & 0.81 $\pm$ 0.18 & (1) \\  \hline 
180620 & 1.118 & 371.90 $\pm$ 49.79 & 8.55 $\pm$ 0.38 & (5) \\  \hline 
000418 & 1.12 & 284.00 $\pm$ 21.00 & 9.51 $\pm$ 1.79 & (1) \\  \hline 
210610 & 1.13 & 665.18 $\pm$ 9.30 & 49.76 $\pm$ 0.01 & (5) \\  \hline 
061126 & 1.159 & 1337.00 $\pm$ 410.00 & 31.42 $\pm$ 3.59 & (1) \\  \hline 
130701A & 1.16 & 191.80 $\pm$ 8.62 & 1.70 $\pm$ 0.50 & (4) \\  \hline 
190324 & 1.172 & 285.96 $\pm$ 7.67 & 9.31 $\pm$ 0.07 & (5) \\  \hline 
140213 & 1.208 & 190.18 $\pm$ 4.10 & 12.56 $\pm$ 0.32 & (5) \\  \hline 
140213A & 1.21 & 176.61 $\pm$ 4.42 & 10.10 $\pm$ 2.60 & (4) \\  \hline 
140907 & 1.21 & 308.19 $\pm$ 10.31 & 2.71 $\pm$ 0.78 & (5) \\  \hline 
130907A & 1.24 & 881.77 $\pm$ 24.62 & 314.00 $\pm$ 79.70 & (4) \\  \hline 
020813 & 1.25 & 590.00 $\pm$ 151.00 & 68.35 $\pm$ 17.09 & (1) \\  \hline 
200829 & 1.25 & 716.24 $\pm$ 4.63 & 124.40 $\pm$ 0.04 & (5) \\  \hline 
061007 & 1.262 & 890.00 $\pm$ 124.00 & 89.96 $\pm$ 8.99 & (1) \\  \hline 
131030A & 1.29 & 405.86 $\pm$ 22.93 & 4.80 $\pm$ 1.50 & (4) \\  \hline 
130420A & 1.3 & 128.63 $\pm$ 6.89 & 7.90 $\pm$ 2.20 & (4) \\  \hline 
990506 & 1.3 & 677.00 $\pm$ 156.00 & 98.13 $\pm$ 9.90 & (1) \\  \hline 
061121 & 1.314 & 1289.00 $\pm$ 153.00 & 23.50 $\pm$ 2.70 & (1) \\  \hline 
141220 & 1.32 & 415.34 $\pm$ 10.07 & 2.72 $\pm$ 0.56 & (5) \\  \hline 
140801 & 1.32 & 276.98 $\pm$ 2.64 & 6.06 $\pm$ 0.19 & (5) \\  \hline 
071117 & 1.331 & 112.00 $\pm$ 56.00 & 5.86 $\pm$ 2.70 & (1) \\  \hline 
070521 & 1.35 & 522.00 $\pm$ 55.00 & 10.81 $\pm$ 1.80 & (3) \\  \hline 
100414 & 1.368 & 1295.00 $\pm$ 120.00 & 54.99 $\pm$ 5.41 & (3) \\  \hline 
120711 & 1.405 & 2340.00 $\pm$ 230.00 & 180.41 $\pm$ 18.04 & (3) \\  \hline 
180205 & 1.409 & 84.80 $\pm$ 17.02 & 0.89 $\pm$ 0.17 & (5) \\  \hline 
100814 & 1.44 & 312.32 $\pm$ 48.80 & 7.70 $\pm$ 3.10 & (4) \\  \hline 
180314 & 1.445 & 251.73 $\pm$ 4.49 & 10.23 $\pm$ 0.68 & (5) \\  \hline 
141221 & 1.452 & 225.87 $\pm$ 28.73 & 2.65 $\pm$ 0.44 & (5) \\  \hline 
110213 & 1.46 & 223.86 $\pm$ 70.11 & 8.80 $\pm$ 4.10 & (4) \\  \hline 
150301 & 1.517 & 460.62 $\pm$ 28.66 & 3.43 $\pm$ 0.59 & (5) \\  \hline 
161117 & 1.549 & 205.62 $\pm$ 3.05 & 23.63 $\pm$ 0.93 & (5) \\  \hline 
110503 & 1.61 & 572.25 $\pm$ 50.95 & 18.90 $\pm$ 5.50 & (4) \\  \hline 
131105 & 1.686 & 721.80 $\pm$ 18.31 & 20.58 $\pm$ 1.71 & (5) \\  \hline 
080928 & 1.692 & 95.00 $\pm$ 23.00 & 3.99 $\pm$ 0.91 & (3) \\  \hline 
100906 & 1.73 & 387.23 $\pm$ 244.07 & 27.70 $\pm$ 11.80 & (4) \\  \hline 
120119 & 1.73 & 417.38 $\pm$ 54.56 & 36.00 $\pm$ 11.70 & (4) \\  \hline 
150314 & 1.758 & 957.48 $\pm$ 7.90 & 89.16 $\pm$ 2.15 & (5) \\  \hline 
110422 & 1.77 & 421.04 $\pm$ 13.85 & 75.80 $\pm$ 16.70 & (4) \\  \hline 
131011 & 1.874 & 625.49 $\pm$ 40.88 & 14.74 $\pm$ 1.59 & (3) \\  \hline 
140623 & 1.92 & 953.53 $\pm$ 138.25 & 3.74 $\pm$ 0.45 & (5) \\  \hline 
060814 & 1.923 & 751.00 $\pm$ 246.00 & 56.71 $\pm$ 5.27 & (1) \\  \hline 
210619 & 1.937 & 799.33 $\pm$ 5.07 & 423.63 $\pm$ 0.12 & (5) \\  \hline 
170113 & 1.968 & 333.92 $\pm$ 58.79 & 2.45 $\pm$ 0.68 & (5) \\  \hline 
170705 & 2.01 & 294.61 $\pm$ 7.64 & 18.31 $\pm$ 0.77 & (5) \\  \hline 
161017 & 2.013 & 718.76 $\pm$ 40.77 & 7.49 $\pm$ 1.55 & (5) \\  \hline 
140620 & 2.04 & 211.21 $\pm$ 10.72 & 9.72 $\pm$ 0.56 & (5) \\  \hline 
081203 & 2.05 & 1541.00 $\pm$ 756.00 & 31.85 $\pm$ 11.83 & (3) \\  \hline 
150403 & 2.06 & 1311.94 $\pm$ 21.06 & 99.28 $\pm$ 2.42 & (5) \\  \hline 
080207 & 2.086 & 333.00 $\pm$ 222.00 & 16.39 $\pm$ 1.82 & (3) \\  \hline 
061222 & 2.088 & 874.00 $\pm$ 150.00 & 30.04 $\pm$ 6.37 & (1) \\  \hline 
130610 & 2.09 & 911.83 $\pm$ 132.65 & 9.00 $\pm$ 3.00 & (4) \\  \hline 
120624 & 2.197 & 1791.00 $\pm$ 134.00 & 282.00 $\pm$ 1.20 & (3) \\  \hline 
121128 & 2.2 & 243.20 $\pm$ 12.80 & 10.40 $\pm$ 3.50 & (4) \\  \hline 
080804 & 2.204 & 810.00 $\pm$ 45.00 & 12.03 $\pm$ 0.55 & (3) \\  \hline 
081221 & 2.26 & 284.00 $\pm$ 14.00 & 31.92 $\pm$ 1.82 & (3) \\  \hline 
130505 & 2.27 & 2063.37 $\pm$ 101.37 & 57.70 $\pm$ 17.90 & (4) \\  \hline 
141028 & 2.33 & 976.02 $\pm$ 17.98 & 76.16 $\pm$ 1.97 & (5) \\  \hline 
131108 & 2.4 & 1247.43 $\pm$ 16.30 & 63.94 $\pm$ 2.57 & (5) \\  \hline 
171222 & 2.409 & 59.80 $\pm$ 4.14 & 3.41 $\pm$ 1.83 & (5) \\  \hline 
190719 & 2.469 & 295.42 $\pm$ 23.25 & 12.17 $\pm$ 0.10 & (5) \\  \hline 
120716 & 2.486 & 397.00 $\pm$ 40.00 & 30.15 $\pm$ 0.27 & (3) \\  \hline 
120811 & 2.671 & 198.00 $\pm$ 19.00 & 6.41 $\pm$ 0.64 & (3) \\  \hline 
140206 & 2.73 & 452.10 $\pm$ 5.83 & 29.69 $\pm$ 3.05 & (5) \\  \hline 
161014 & 2.823 & 646.18 $\pm$ 14.42 & 9.62 $\pm$ 1.05 & (5) \\  \hline 
181020 & 2.938 & 1544.13 $\pm$ 28.97 & 80.25 $\pm$ 0.07 & (5) \\  \hline 
060607 & 3.075 & 478.00 $\pm$ 118.00 & 11.93 $\pm$ 2.75 & (1) \\  \hline 
140423 & 3.26 & 494.90 $\pm$ 15.89 & 69.38 $\pm$ 2.61 & (5) \\  \hline 
140808 & 3.29 & 503.85 $\pm$ 6.46 & 8.99 $\pm$ 0.63 & (5) \\  \hline 
110818 & 3.36 & 1117.47 $\pm$ 241.11 & 25.60 $\pm$ 8.50 & (4) \\  \hline 
060306 & 3.5 & 315.00 $\pm$ 135.00 & 7.63 $\pm$ 1.01 & (3) \\  \hline 
151111 & 3.5 & 533.91 $\pm$ 50.33 & 5.43 $\pm$ 1.84 & (5) \\  \hline 
170405 & 3.51 & 1204.23 $\pm$ 9.29 & 255.20 $\pm$ 5.02 & (5) \\  \hline 
100704 & 3.6 & 809.60 $\pm$ 135.70 & 19.06 $\pm$ 1.91 & (4) \\  \hline 
130408 & 3.76 & 1003.94 $\pm$ 137.98 & 28.90 $\pm$ 9.60 & (4) \\  \hline 
060210 & 3.91 & 574.00 $\pm$ 187.00 & 32.23 $\pm$ 1.84 & (3) \\  \hline 
120712 & 4.174 & 641.00 $\pm$ 130.00 & 21.19 $\pm$ 1.84 & (3) \\  \hline 
130606 & 5.91 & 2031.54 $\pm$ 483.70 & 28.60 $\pm$ 11.60 & (4) \\  \hline 
050318 & 1.44 & 115.00 $\pm$ 25.00 & 2.34 $\pm$ 0.17 & (1) \\  \hline 
010222 & 1.48 & 766.00 $\pm$ 30.00 & 85.57 $\pm$ 8.79 & (1) \\  \hline 
120724 & 1.48 & 68.45 $\pm$ 18.60 & 0.88 $\pm$ 0.12 & (4) \\  \hline 
060418 & 1.489 & 572.00 $\pm$ 143.00 & 13.63 $\pm$ 2.96 & (1) \\  \hline 
030328 & 1.52 & 328.00 $\pm$ 55.00 & 39.42 $\pm$ 3.69 & (1) \\  \hline 
070125 & 1.547 & 934.00 $\pm$ 148.00 & 84.62 $\pm$ 8.27 & (1) \\  \hline 
090102 & 1.547 & 1149.00 $\pm$ 166.00 & 22.14 $\pm$ 4.01 & (2) \\  \hline 
040912 & 1.563 & 44.00 $\pm$ 33.00 & 1.36 $\pm$ 0.39 & (1) \\  \hline 
990123 & 1.6 & 1724.00 $\pm$ 466.00 & 242.38 $\pm$ 39.27 & (1) \\  \hline 
071003 & 1.604 & 2077.00 $\pm$ 286.00 & 36.18 $\pm$ 4.01 & (2) \\  \hline 
090418 & 1.608 & 1567.00 $\pm$ 384.00 & 16.06 $\pm$ 4.03 & (2) \\  \hline 
990510 & 1.619 & 423.00 $\pm$ 42.00 & 17.99 $\pm$ 2.77 & (1) \\  \hline 
080605 & 1.64 & 650.00 $\pm$ 55.00 & 24.08 $\pm$ 1.98 & (2) \\  \hline 
131105A & 1.686 & 547.68 $\pm$ 83.53 & 35.39 $\pm$ 1.19 & (4) \\  \hline 
091020 & 1.71 & 507.23 $\pm$ 68.20 & 8.40 $\pm$ 1.08 & (3) \\  \hline 
120326 & 1.798 & 129.97 $\pm$ 10.27 & 3.68 $\pm$ 0.17 & (4) \\  \hline 
080514B & 1.8 & 627.00 $\pm$ 65.00 & 17.01 $\pm$ 4.03 & (2) \\  \hline 
090902B & 1.822 & 2187.00 $\pm$ 31.00 & 277.68 $\pm$ 8.66 & (4) \\  \hline 
020127 & 1.9 & 290.00 $\pm$ 100.00 & 3.51 $\pm$ 0.09 & (1) \\  \hline 
080319C & 1.95 & 906.00 $\pm$ 272.00 & 14.53 $\pm$ 2.91 & (1) \\  \hline 
081008 & 1.968 & 261.00 $\pm$ 52.00 & 9.45 $\pm$ 0.89 & (2) \\  \hline 
030226 & 1.98 & 289.00 $\pm$ 66.00 & 12.94 $\pm$ 0.99 & (1) \\  \hline 
130612 & 2.006 & 186.07 $\pm$ 31.56 & 0.81 $\pm$ 0.10 & (4) \\  \hline 
000926 & 2.07 & 310.00 $\pm$ 20.00 & 27.98 $\pm$ 6.46 & (1) \\  \hline 
090926 & 2.106 & 974.00 $\pm$ 50.00 & 167.34 $\pm$ 8.54 & (3) \\  \hline 
011211 & 2.14 & 186.00 $\pm$ 24.00 & 5.71 $\pm$ 0.68 & (1) \\  \hline 
071020 & 2.145 & 1013.00 $\pm$ 160.00 & 9.97 $\pm$ 4.58 & (1) \\  \hline 
050922C & 2.198 & 415.00 $\pm$ 111.00 & 5.62 $\pm$ 1.91 & (1) \\  \hline 
110205 & 2.22 & 740.60 $\pm$ 322.00 & 40.39 $\pm$ 8.27 & (4) \\  \hline 
060124 & 2.296 & 784.00 $\pm$ 285.00 & 43.85 $\pm$ 6.45 & (1) \\  \hline 
021004 & 2.3 & 266.00 $\pm$ 117.00 & 3.49 $\pm$ 0.52 & (1) \\  \hline 
051109A & 2.346 & 539.00 $\pm$ 200.00 & 6.83 $\pm$ 0.67 & (1) \\  \hline 
060908 & 2.43 & 514.00 $\pm$ 102.00 & 10.38 $\pm$ 0.99 & (1) \\  \hline 
080413 & 2.433 & 584.00 $\pm$ 180.00 & 7.98 $\pm$ 1.99 & (2) \\  \hline 
090812 & 2.452 & 2000.00 $\pm$ 700.00 & 44.43 $\pm$ 7.65 & (4) \\  \hline 
100728B & 2.453 & 359.11 $\pm$ 48.34 & 4.19 $\pm$ 0.14 & (4) \\  \hline 
130518 & 2.49 & 1382.04 $\pm$ 31.41 & 182.93 $\pm$ 1.19 & (4) \\  \hline 
081121 & 2.512 & 871.00 $\pm$ 123.00 & 25.73 $\pm$ 4.97 & (2) \\  \hline 
081118 & 2.58 & 147.00 $\pm$ 14.00 & 4.25 $\pm$ 0.89 & (2) \\  \hline 
080721 & 2.591 & 1741.00 $\pm$ 227.00 & 124.66 $\pm$ 21.73 & (2) \\  \hline 
050820 & 2.612 & 1325.00 $\pm$ 277.00 & 102.89 $\pm$ 8.04 & (1) \\  \hline 
030429 & 2.65 & 128.00 $\pm$ 26.00 & 2.31 $\pm$ 0.33 & (1) \\  \hline 
120811C & 2.671 & 157.49 $\pm$ 20.92 & 12.35 $\pm$ 1.17 & (4) \\  \hline 
080603B & 2.69 & 376.00 $\pm$ 100.00 & 10.81 $\pm$ 0.98 & (2) \\  \hline 
140206A & 2.73 & 447.60 $\pm$ 22.38 & 29.27 $\pm$ 0.52 & (4) \\  \hline 
091029 & 2.752 & 230.00 $\pm$ 66.00 & 8.25 $\pm$ 0.77 & (3) \\  \hline 
081222 & 2.77 & 505.00 $\pm$ 34.00 & 29.64 $\pm$ 3.02 & (2) \\  \hline 
050603 & 2.821 & 1333.00 $\pm$ 107.00 & 64.03 $\pm$ 3.66 & (1) \\  \hline 
110731 & 2.83 & 1164.32 $\pm$ 49.79 & 46.16 $\pm$ 0.18 & (4) \\  \hline 
111107 & 2.893 & 420.44 $\pm$ 124.58 & 3.43 $\pm$ 0.57 & (4) \\  \hline 
050401 & 2.9 & 467.00 $\pm$ 110.00 & 36.39 $\pm$ 7.66 & (1) \\  \hline 
090715B & 3.0 & 536.00 $\pm$ 172.00 & 22.08 $\pm$ 3.44 & (4) \\  \hline 
080607 & 3.036 & 1691.00 $\pm$ 226.00 & 185.12 $\pm$ 9.92 & (2) \\  \hline 
081028 & 3.038 & 234.00 $\pm$ 93.00 & 16.75 $\pm$ 1.96 & (2) \\  \hline 
120922 & 3.1 & 156.62 $\pm$ 0.04 & 33.99 $\pm$ 3.85 & (4) \\  \hline 
020124 & 3.2 & 448.00 $\pm$ 148.00 & 27.02 $\pm$ 2.25 & (1) \\  \hline 
060526 & 3.21 & 105.00 $\pm$ 21.00 & 2.72 $\pm$ 1.36 & (1) \\  \hline 
080810 & 3.35 & 1470.00 $\pm$ 180.00 & 44.15 $\pm$ 4.85 & (2) \\  \hline 
030323 & 3.37 & 270.00 $\pm$ 113.00 & 2.94 $\pm$ 0.98 & (1) \\  \hline 
971214 & 3.42 & 685.00 $\pm$ 133.00 & 22.06 $\pm$ 2.76 & (1) \\  \hline 
060707 & 3.425 & 279.00 $\pm$ 28.00 & 5.78 $\pm$ 1.01 & (1) \\  \hline 
060115 & 3.53 & 285.00 $\pm$ 34.00 & 6.59 $\pm$ 1.06 & (1) \\  \hline 
090323 & 3.57 & 1901.00 $\pm$ 343.00 & 402.48 $\pm$ 49.17 & (3) \\  \hline 
130514 & 3.6 & 496.80 $\pm$ 151.80 & 51.19 $\pm$ 6.81 & (4) \\  \hline 
120802 & 3.796 & 274.33 $\pm$ 93.04 & 12.74 $\pm$ 2.07 & (4) \\  \hline 
100413 & 3.9 & 1783.60 $\pm$ 374.85 & 72.95 $\pm$ 23.80 & (4) \\  \hline 
120909 & 3.93 & 1651.55 $\pm$ 123.25 & 84.16 $\pm$ 7.19 & (4) \\  \hline 
131117A & 4.042 & 221.85 $\pm$ 37.31 & 1.63 $\pm$ 0.33 & (4) \\  \hline 
060206 & 4.048 & 394.00 $\pm$ 46.00 & 4.59 $\pm$ 0.98 & (1) \\  \hline 
090516 & 4.109 & 971.00 $\pm$ 390.00 & 65.78 $\pm$ 12.75 & (4) \\  \hline 
080916C & 4.35 & 2646.00 $\pm$ 566.00 & 371.24 $\pm$ 78.06 & (2) \\  \hline 
000131 & 4.5 & 987.00 $\pm$ 416.00 & 181.48 $\pm$ 30.89 & (1) \\  \hline 
111008 & 5.0 & 894.00 $\pm$ 240.00 & 48.05 $\pm$ 4.99 & (4) \\  \hline 
060927 & 5.6 & 475.00 $\pm$ 47.00 & 14.49 $\pm$ 2.15 & (1) \\  \hline 
050904 & 6.29 & 3178.00 $\pm$ 1094.00 & 127.35 $\pm$ 12.74 & (1) \\  \hline 
080913 & 6.695 & 710.00 $\pm$ 350.00 & 8.36 $\pm$ 2.44 & (2) \\  \hline 
090423 & 8.2 & 491.00 $\pm$ 200.00 & 11.15 $\pm$ 2.97 & (2) \\  \hline 
\hline
\caption{\label{GRBsample} 221 GRBs with redshifts, peak energy in cosmological rest frame and isotropic-equivalent energy. The 1\,$\sigma$ uncertainties are also given.\\
(a) $E_{\rm iso}$ is computed with cosmological parameters: $H_{0}$=67.4 $\mathrm{kms}^{-1} \mathrm{Mpc}^{-1}$, $\Omega_{\mathrm{M}}=0.315$, $ \Omega_{\Lambda}=0.685$. \\
(b) References for GRBs: (1)\citet{2008MNRAS.391..577A};(2)\citet{2009A&A...508..173A} ;(3)\citet{2019MNRAS.486L..46A}; (4)\citet{2016A&A...585A..68W};(5)\citet{2020ApJ...893...46V,2014ApJS..211...12G,2014ApJS..211...13V,2016ApJS..223...28N}} \centering
\end{longtable}
}

\clearpage

	\begin{table*}
	\caption{ The $E_{\mathrm{iso }}-E_{\mathrm{p}}$ correlation fitting results in five redshift bins. We give the best-fit values with 1$\sigma$ uncertainties. The first column is the redshfit range of each bin. The last column is the number of GRBs in redshift bins.}\label{T1}
	\centering
	\begin{tabular}{|c |c  c  c |c|} 
		\hline\hline
		Redshift range  & $a$             & $b$             & $\sigma_{\mathrm{ext}}$       &  Number of GRBs \\ \hline\hline
		[0,0.55]     & 48.83 $\pm$ 0.34   & 1.56 $\pm$ 0.16  & 0.41 $\pm$ 0.09   & 20 \\ \hline
		[0.55,1.18]  &  49.11 $\pm$ 0.35  &   1.47 $\pm$ 0.14       &  0.38 $\pm$ 0.04      & 54 \\ \hline
		[1.18,1.74]  &  50.01 $\pm$ 0.46     &   1.21 $\pm$ 0.17      &  0.42 $\pm$ 0.05      & 44 \\ \hline
		[1.74,2.55]  &  49.91 $\pm$ 0.53     &   1.25 $\pm$ 0.19      &  0.43 $\pm$ 0.05      & 48 \\ \hline
		[2.55,8.20]   &  49.74 $\pm$ 0.38     &   1.30 $\pm$ 0.13      &  0.34 $\pm$ 0.04      & 55\\ \hline
		\hline
	\end{tabular}	
\end{table*}

	\begin{table*}
	\caption{ The best-fit results of six sub-samples. The number of each sub-samples are given in the last column.}		\label{T2}
	\centering
	\begin{tabular}{|c c|c  c  c |c|} 
		\hline\hline
		$z_{min}$	& $z_{max}$ & $a$             & $b$             & $\sigma_{\mathrm{ext}}$       &  Number of GRBs \\ \hline\hline
		0.736   & 0.807 & 49.57 $\pm$ 0.87    & 1.37 $\pm$ 0.36  & 0.32 $\pm$ 0.12      & 6 \\ \hline
		0.897  & 1.092 & 49.06 $\pm$ 0.75    & 1.51 $\pm$ 0.28  & 0.36 $\pm$ 0.08      & 14 \\ \hline
		1.100     & 1.210  & 48.81 $\pm$ 1.02    & 1.64 $\pm$ 0.41  & 0.35 $\pm$ 0.09      & 11 \\ \hline
		1.350    & 1.489 & 49.14 $\pm$ 0.45    & 1.51 $\pm$ 0.17  & 0.24 $\pm$ 0.08      & 12 \\ \hline
		2.469   & 2.671 & 49.35 $\pm$ 0.48    & 1.50 $\pm$ 0.17  & 0.19 $\pm$ 0.07      & 9\\ \hline
		2.612   & 2.770  & 49.67 $\pm$ 0.70    & 1.40 $\pm$ 0.28  & 0.20 $\pm$ 0.09      & 8\\ \hline
		\hline
	\end{tabular}	
\end{table*}
%

\bsp	
\label{lastpage}
\end{document}